\documentclass[aps,pre,groupedaddress,twocolumn,showpacs]{revtex4}

\usepackage[pdftex]{graphicx}
\usepackage{epstopdf}
\usepackage{amsmath}

\begin{document}

%\preprint{}

\title{Entropy production and Kullback-Leibler divergence between stationary trajectories of discrete systems}
%\title{Entropy production 


\author{\'Edgar Rold\'an}
\author{Juan M.R. Parrondo}
%\thanks{}
\affiliation{Departamento de F\'{\i}sica At\'omica, Molecular y Nuclear and GISC.
Universidad Complutense de Madrid. 28040-Madrid, Spain}


\date{\today}

\begin{abstract}
The irreversibility of a stationary time series can be quantified using the Kullback-Leibler divergence (KLD) between the probability to observe the series and the probability to observe the time-reversed series. Moreover, this KLD is a tool to estimate entropy production from stationary trajectories since it gives a lower bound to the entropy production of the physical process generating the series.  In this paper we introduce analytical and numerical techniques to estimate the KLD between time series generated by several stochastic dynamics with a finite number of states. We examine the accuracy of our estimators for a specific example, a discrete flashing ratchet, and investigate how close is the KLD to the entropy production depending on the number of degrees of freedom of the system that are sampled in the trajectories.
\end{abstract}

%\keywords{}

\maketitle

\section{Introduction}

The relationship between irreversibility and entropy production is
mentioned in many undergraduate courses of thermodynamics and
statistical physics. A canonical example is a glass falling to the ground and smashing into pieces.
The time-reverse of this process is compatible with Newton's laws, but the chances for it  to occur
spontaneously are incredibly small. Such a process is irreversible
and the signature of this irreversibility is the production
of a macroscopic amount of entropy in the universe. 

The relation between irreversibility and entropy production was only a qualitative statement until the recent introduction of the Kullback-Leibler divergence (KLD) in the context of fluctuation theorems \cite{kpb,njp}. The time irreversibility of a process is given by the distinguishability between the process and its time reversal, which in turn can be quantified using the KLD or relative entropy, a measure of the distinguishability between two probability distributions defined in information theory \cite{cover,njp,crooks2011}. This KLD, multiplied by the Boltzmann constant, turns out to be a lower bound to the entropy production along the process 
\cite{kpb,njp,maes,jarz2006,gaspard_markov,zolfaghari,roldan,lebowitz,mazonka,vanZon,seifert,ciliberto,jordan}.
The bound becomes  more accurate when the observables  that are used to calculate the KLD contain a more complete description of the state of the system. This result has been
derived in a variety of situations such as driven systems under Hamiltonian~\cite{kpb,njp,zolfaghari} and Langevin~\cite{footprints,ciliberto,
kurchan, mazonka} dynamics; Markovian processes~\cite{gaspard_markov, lebowitz}; and also for 
electrical circuits~\cite{vanZon}. Andrieux {\em et al.} have verified it  experimentally using the data of the position of a Brownian particle in a
moving optical trap~\cite{ciliberto,garnier}, and we have shown that the bound yields useful estimates of  the entropy production in non equilibrium 
stationary states (NESS)~\cite{roldan}.

Imagine repeatedly sampling (or measuring) an observable of a system in a NESS. The trajectory of the outcomes is a stationary time series that can be used to estimate the KLD, by comparing the statistics of the time series with the statistics of the same series but time reversed~\cite{roldan}. This means that one can bound from below the entropy production in the NESS from a single time series obtained in an experiment. Such a tool is of interest in many practical situations. For instance, it allows one to discriminate between active and passive processes in biological systems, or even to estimate or bound the amount of entropy produced, and therefore the amount of ATP consumed in a biological process. In fact, there have been previous attempts  to
make this distinction. Martin {\em et al.}  have considered the violation of the fluctuation-dissipation relationship as a signature of non-equilibrium in the motion of a hair cell by using two types of measurement: the spontaneous motion of the hair bundle and the response to an external force~\cite{julicher}. Amman {\em et al.} discriminated between equilibrium and NESS in a three state chemical system~\cite{seifert2}. Finally, Kennel introduced in~\cite{kennel}
criteria based on compression algorithms to distinguish between time symmetric and time asymmetric chaotic series but without any connection to the physical entropy.

We are interested in estimating the KLD between the probability of observing a stationary trajectory of one or several observables of the system and the probability of observing the same trajectory but time reversed. We want to explore how this quantity bounds the entropy production of the underlying physical process~\cite{kpb,njp,roldan} depending on the  number of degrees of freedom of the system that are sampled in the observed stationary trajectory. Two distinct issues immediately arise: 
 the estimation of the KLD from an empirical stationary time series and the accuracy of the bound. In this paper we address these two issues by introducing numerical and semi-analytical techniques to estimate the KLD from data obtained from systems with a finite number of states. 
 
 There have been different attempts to provide accurate estimators of the KLD from a finite number of data.
References~\cite{cou,rached} investigate how this measure can be estimated when considering empirical
probability distributions of two different Markovian and higher order Markovian time series. They develop
techniques based on empirical counting of finite sequences of data which are generalized to real-valued time 
series in Refs.~\cite{wang, budka, ciliberto}. A different approach is given in~\cite{ziv}, where the KLD between two 
different probability distributions is estimated using compression algorithms.
In this paper we refine these methods and test their performance when used to estimate the KLD from single stationary trajectories.

To explore the bound to the entropy production, we work with a 
discrete flashing ratchet model, where we can compare the entropy production with the analytical value and the empirical estimations of the KLD.
With this model, we can analyze how information losses affect the estimation of the KLD and the tightness of the bound for the entropy production. 

The  paper is organized as follows: section \ref{sec:kld} reviews the concept of the KLD, and discusses its connection with entropy production. In Section \ref{sec:hmc} we present novel analytical and semi-analytical tools to calculate the KLD between hidden Markov chains. Section \ref{sec:estimators} gives a detailed description of the estimators of the KLD from empirical data, whose performance for the flashing ratchet is analyzed in Sec. \ref{sec:flash}. Finally, we present our main conclusions in Sec. \ref{sec:concl}.

\section{Kullback-Leibler divergence, irreversibility, and entropy production}
\label{sec:kld}

\subsection{The Kullback-Leibler divergence}

The Kullback-Leibler divergence, or relative entropy, measures
the distinguishability  of two probability distributions $p(x)$ and
$q(x)$:
\begin{equation}
 D[p(x)||q(x)]= \int dx\, p(x) \log\frac{p(x)}{q(x)}.
\label{kld}
\end{equation}
It is always positive, and vanishes if and only if $p(x)=q(x)$ for all $x$. Its interpretation as a measure of distinguishability is a consequence of the Chernoff-Stein lemma~\cite{cover}: the probability of incorrectly guessing (via hypothesis testing) that a sequence of $n$ data is distributed according to $p$ when the true distribution is $q$ is asymptotically equal to $e^{-nD[p(x)||q(x)]}$.  Therefore, when $p$ and $q$ are similar $-$in the sense that they overlap significantly$-$ the likelihood of incorrectly guessing the distribution, $p$ or $q$, is large~\cite{cover}.

Let us recall a property of the KLD that we will use throughout the paper~\cite{cover}. If we have two random variables $X,Y$ and two 
joint probability distributions $p(x,y)$ and $q(x,y)$, then
\begin{equation}
\label{eq:chain}
D[p(x,y)||q(x,y)]\geq  D[p(x)||q(x)].
\end{equation}
This means that it is harder to distinguish between $p$ and $q$ when we consider only
the marginal distributions, $p(x)$ and $q(x)$, instead of the full
joint distributions, $p(x,y)$ and $q(x,y)$.
If $X,Y$ describe the state of a system, Eq.~\eqref{eq:chain} indicates that 
the KLD decreases when only a partial description of the system, given by the variable $X$, is available.

\subsection{Irreversibility and entropy production}
\label{sec:irrev}

Consider a physical system with Hamiltonian $H(z;\lambda)$, where $z$ denotes a point in
 phase space $\Gamma$, and $\lambda$ is a parameter of the system controlled by
an external agent. The system is initially isolated in equilibrium
at temperature $T$, and the external agent modifies $\lambda$
following a protocol $\lambda_{t}$, with $t\in [0,\tau]$. 
 We then
let the system equilibrate by coupling it to a bath at temperature
$T'$. The initial and final states of this process are equilibrium
states for which entropy is well defined. 
We denote by $\rho(z,t)$ the probability density on phase space at time $t$,
 and by $\tilde\rho(\tilde z,t)$  the probability density when the system is driven by the time-reversed protocol
 $\tilde\lambda_{t}=\lambda_{\tau-t}$ with $t\in [0,\tau]$. Here $\tilde z$ denotes the point in phase space resulting from changing the sign of all
momenta in $z$.  In Ref.~\cite{njp} it is
proved that the change of the entropy $\Delta S$ in the system plus the
bath, averaged over many realizations of the process, satisfies
\begin{equation}\label{eq:main}
\langle \Delta S\rangle =kD[\rho(z,t)||\tilde\rho(\tilde z,\tau-t)],
\end{equation}
where $k$ is Boltzmann's constant. Equation \eqref{eq:main} is valid for a variety of initial
equilibrium conditions~\cite{njp}: canonical, multi-canonical
(several uncoupled systems at different temperatures), and
grand-canonical distributions, as well as for different types of
baths equilibrating the system at the end of the  process. In particular, for canonical initial conditions in the forward and in the time-reversed processes, both at the same temperature $T$, Eq.~\eqref{eq:main} reads (see Ref.~\cite{njp})

\begin{eqnarray}\langle \Delta
S\rangle &=& \langle\Delta S_{\rm system}\rangle+\langle\Delta
S_{\rm bath}\rangle
\nonumber \\
&=& \frac{  \langle \Delta E\rangle -\Delta F}{T} +\frac{\langle
Q\rangle }{T}
\nonumber \\
&=& \frac{ \langle  W\rangle-\Delta F}{T},
\end{eqnarray}
where $\langle \Delta E\rangle $ and $\Delta F$ refer respectively
to the system average energy and free energy change, $Q$ is the heat
exchanged with the thermal bath at the end of the process (realization dependent), and $W=\Delta E+Q$ is the work performed by
the external agent. Therefore, in this specific case, entropy
production equals the average dissipated work $\langle W_{\rm
diss}\rangle = \langle W\rangle -\Delta F$ divided by the
temperature $T$ and \eqref{eq:main} becomes
\begin{equation}\label{eq:main2}
\langle W_{\rm diss}\rangle =kTD[\rho(z,t)||\tilde\rho(\tilde z,\tau-t)].
\end{equation}
Since the evolution is deterministic, except for the last stage
where the system is connected to the bath, the point $z$ at time $t$
determines the whole trajectory of the system $\{z(t)\}_{t=0}^\tau$.
Then $z(t)$ and $\{z(t)\}_{t=0}^\tau$ carry the same information and
the KLD of their respective PDF's are equal. Equation~\eqref{eq:main2} can be rewritten in terms of path probabilities $\mathcal P$~\cite{footprints} 
\begin{equation}\label{eq:main3}
\langle W_{\rm diss}\rangle=kTD[\mathcal P (\{z(t)\}_{t=0}^\tau  )|| \widetilde\mathcal P ( \{ \tilde z(\tau-t)\} _{t=0}^\tau )].
\end{equation}
On the other hand,  integrating Crook's relationship~\cite{crooks}, $W-\Delta F=\log \frac{p(W)}{\tilde p(-W)}$, where $p(W)$ [$\tilde p(W)$] is the probability density of the work done on the system along the actual (time-reversed) process~\cite{crooks,footprints}, one immediately gets
\begin{equation}
\label{eq:crooks}
\langle W_{\rm diss}\rangle = kTD[p(W)||\tilde p(-W)].
\end{equation}
Notice that the work $W$ is a function of the trajectory $\{z(t)\}_{t=0}^\tau$ containing much less information than the trajectory itself. As indicated by Eq.~\eqref{eq:chain}, the KLD of work distributions should in principle be smaller than the KLD of trajectory distributions.
On the contrary, the KLD is the same, indicating that all the irreversibility of the process is captured by the dissipative work~\cite{footprints}.

\subsection{Stationary trajectories}

We now proceed to apply the above results to stationary
trajectories. Consider a long process in which the system reaches a
non-equilibrium stationary state (NESS) after a possible initial
transient. In the NESS the external parameter is held fixed, $\lambda_t=\lambda$; 
the system is kept out of equilibrium due to the existence of baths
at different temperatures (a possibility that is included in the
hypothesis used in \cite{njp} to prove \eqref{eq:main}) or different
chemical potentials, external constant forces, etc. In the steady state, since the control parameter remains fixed, the protocol and its time reversal are identical $\lambda_t=\tilde\lambda_t=\lambda$~\cite{seifert}. Therefore the probability distributions of the process and its time reversal are identical, $\mathcal{\widetilde P}=\mathcal{P}$.  In the long time limit, $\tau\to\infty$, we can neglect the contribution of the transient to the entropy production and rewrite \eqref{eq:main} for the entropy production per unit of time $\dot{S}$ in the NESS~\cite{maes2} as

\begin{equation}
\langle \dot {S}\rangle =
\lim_{\tau\to\infty}\frac{k}{\tau} D\left[\mathcal{P}\left(\left\{z(t)\right\}_{t=0}^\tau\right) \right|\left|\mathcal{P}\left(\left\{\tilde z(\tau-t)\right\}_{t=0}^\tau\right)\right].
\label{diss_ness}
\end{equation} 
A similar expression can be obtained from the Gallavotti-Cohen
theorem~\cite{cohen,koen}, $\Delta S \simeq k\log\frac{p_\tau(\Delta S)}{p_\tau(-\Delta S)}$, where $p_\tau(\Delta S)$ is the
probability to observe an entropy production $\Delta S$ in the
interval $[0,\tau]$. The Gallavotti-Cohen relationship, which is exact for $\tau\to\infty$, yields, after averaging
\begin{equation}
\label{eq:gallavotti}
\langle \dot {S}\rangle =\lim_{\tau\to\infty}\frac{k}{\tau} D\left[p_\tau(\Delta S) || p_\tau(-\Delta S)\right].
\end{equation}
Consequently, although $\Delta S$ is another observable that is obtained as a function of the microstate of the system, the KLD calculated with $\Delta S$ yields the same value as the one calculated with full information of the system. Therefore entropy production captures all the information about the time irreversibility of the NESS.

%Analogously to our previous discussion of Eqs.~\eqref{eq:main3} and
%\eqref{eq:crooks}, entropy production $\Delta s$ is another observable that is obtained as a function of the microstate of the system and saturates the bound 
%
%the  entropy production is a functional of the
%trajectory containing less information than the microscopic trajectory.
%Nevertheless, the KLD does not decrease, indicating that the entropy production
% captures all the information about the time irreversibility of the NESS.

When one does not observe the entire microscopic trajectory $\left\{z(t)\right\}_{t=0}^\tau$ in \eqref{diss_ness}  but the trajectory followed by one or several observables of the system $x(t)$, the KLD only provides a lower bound to the entropy production~\cite{alexpre}. Equations~\eqref{eq:crooks} and~\eqref{eq:gallavotti} indicate that the equality is recovered if the observables determine in a unique way the entropy production or the dissipated work.

In an experimental context, the observables are usually sampled at a finite frequency. 
The output is then a time series of data or discrete {\em trajectory}, $\mathbf{x} =  (\hat x_1,\hat x_2,\cdots, \hat x_n)$, where
$\hat x_i$ can be the value of a single or several observables of the system. In this case, we are interested
in estimating the entropy production \emph{per data} of the underlying physical process, which we
denote by $\langle\dot{S}\rangle$ in the rest of the paper. Entropy production per data is related to the KLD rate {\em per data}, 
which we define below.

Given an infinitely long realization or time series sampled from a random process $X_i$ ($i=1,2,\dots$), which can be multi-dimensional, we define by 
$p(x_1^m)$ the probability that a given string of $m$ consecutive data is equal to $x_1^m=(x_1,x_2,\cdots,x_m)$. We define the $m-$th order KLD for this
 random process $X_i$ by the distinguishability between $p(x_1^m)$ and the probability $p(x_m^1)$ to observe the reverse sequence of data $x_m^1=(x_m,x_{m-1},\cdots,x_1)$.
 
\begin{equation}
 D_m^{X}=D[p(x_1^m) || p(x_m^1)]= \sum_{x_1,\cdots ,x_m} p(x_1^m) \log\frac{p(x_1^m)}{p(x_m^1)}.
\label{Dn}
\end{equation}
The KLD rate for the process $X_i$ is defined as the growth rate of $D_m^{X}$ with the number of data,
\begin{equation}
 d^{X} = \lim_{m\to\infty} \frac{D_m^{X}}{m}.
\label{d}
\end{equation}
By virtue of \eqref{diss_ness} and \eqref{eq:chain}, this quantity bounds from below the entropy production per data
\begin{equation}
 \langle \dot{S}\rangle \geq k\, d^{X},
 \label{main_equation}
\end{equation}
where the bound is saturated if the random variable is the microstate of the system $X=\{\mathbf{q},\mathbf{p}\}$ and the sampling rate is infinite~\cite{alexpre} or $X$ determines uniquely the entropy production in the process.

Equation (\ref{main_equation})  is our basic result. It reveals a striking connection
between physics and the statistics of a time series. 
The left-hand side, $\langle\dot S\rangle$, is a purely physical quantity, whereas the right-hand side, $d^{X}$, is a statistical magnitude depending solely on the observed data, but not
on the physical mechanism generating the data. Such a connection generalizes Landauer's principle relating entropy production and logical irreversibility in computing
machines~\cite{kpb,landauer,gasp}. Equation (\ref{main_equation}) extends this principle and suggests that we can determine the average dissipation of an arbitrary NESS,
even ignoring any physical detail of the system. 

%As mentioned in the Introduction, we could, for instance, determine whether a biological process is active or
%passive or even estimate, or bound, the amount of consumed ATP by measuring the KLD of data generated in the process.

\subsection{Markovian trajectories obeying local detailed balance}
\label{sec:markov}

We first analyze how the bound~\eqref{main_equation} is expressed for Markovian time series that obey detailed
balance by deriving analytical expressions for both entropy production and the KLD rate. If the random
process $X_i$ is Markovian, the probability distribution $p(x_1^m)$ factorizes $p(x_1^m)=p(x_{1})p(x_{2}|x_{1})\cdots
p(x_{m}|x_{m-1})$, which also holds if we reverse the arguments, i.e., for $p(x_m^1)$. Substituting these expressions into equation (\ref{d}), we get
\begin{equation}
d^{X}=\sum_{x_{1},x_{2}}p(x_{1},x_{2})\log\frac{p(x_{2}|x_{1})}{p(x_{1}|x_{2})}=D_{2}^{X}-D_{1}^{X}=D_{2}^{X},
\label{d2}
\end{equation}
since $D_{1}^{X}=0$ when comparing a trajectory and its reverse. Therefore, $d^{X}$ only depends on transition probabilities if  $X$ is a random Markovian process.

We now relate $d^{X}$ in Eq.~\eqref{d2} with the entropy production when the system reaches a NESS, because it is in contact with several thermal baths. In this situation, the local detailed balance condition is satisfied. We call $V(x_i)$ is the energy of the state $x_i$, and $T_{x_1,x_2}$ is the temperature of the bath that activates the transitions $x_1\to x_2$ and $x_2\to x_1$.
%A system obeying detailed balance relaxes to equilibrium and does not reach a NESS. 
The local detailed balance condition  reads in this case
\begin{equation}
\frac{p(x_2|x_1)}{p(x_1|x_2)}=\exp\left(\frac{V(x_1)-V(x_2)}{k\, T_{x_1,x_2}}\right).
\label{db}
\end{equation}
 Inserting \eqref{db} into \eqref{d2},
 \begin{eqnarray}
d^{X} &=& \sum_{x_{1},x_{2}}p(x_{1},x_{2})\frac{V(x_{1})-V(x_{2})}{k\,T_{x_{1},x_{2}}}\nonumber
\\ &=& \sum_{x_{1},x_{2}}p(x_{1},x_{2})\frac{ Q_{x_{1},x_{2}} }{k\, T_{x_{1},x_{2}}}=\frac{\langle \dot S\rangle}{k},
\label{markovequality}
\end{eqnarray}
where $Q_{x_{1},x_{2}}=V(x_{1})-V(x_{2})$ is the heat dissipated to the corresponding thermal bath in the jump $x_1\to x_2$, and $\dot
S$ is the total entropy production per data.  Therefore, Eq.~\eqref{main_equation} is reproduced, with equality, in the case of a physical system obeying local detailed balance, if we have access to all the variables describing the system. The same conclusion is reached if we induce the NESS by means of non-conservative constant forces.

Equation~\eqref{d2} can be explored further by defining the {\em current} 
 from the state $x_1$ to the state $x_2$ as the net probability flow from $x_1$ to $x_2$, $J_{x_1\to x_2}= p(x_1,x_2)-p(x_2,x_1)$. 
 If the system is not far from equilibrium the current tends to zero, and the following condition is satisfied ~$J_{x_1\to x_2}\ll p(x_1,x_2)$, yielding
\begin{equation}
\frac{\langle\dot  S\rangle}{k} = d^{X}= D_{2}^{X} \simeq \sum_{x_1,x_2} \frac{(J_{x_1\to x_2})^{2}}{2p(x_1,x_2)}.
\label{D2J}
\end{equation}
This expression is well known from linear irreversible thermodynamics~\cite{ratchet}, where entropy production is given by
the product of a flow times a thermodynamic force that is proportional to the flow itself. Equation~\eqref{D2J}  implies that the
time asymmetry of a Markovian process not far from equilibrium is revealed by the currents or probability flows that can be observed. In other words, a Markovian process without
flows is time reversible. This is not the case for non-Markovian time series, where irreversibility can show up even in the absence of currents (see below and~\cite{roldan}).

\section{Kullback-Leibler divergence between hidden Markov chains}
\label{sec:hmc}

In many experimental situations, a physical process is Markovian at a micro- or mesoscopic level of description, but the observed time series only contain a subset of the relevant observables, being non-Markovian in general. This is the case in biological systems, where one can only register the behavior of some mechanical and maybe a few chemical variables, while most of the relevant chemical variables cannot be monitored. These kind of non-Markovian time series obtained from an underlying Markov process are called \emph{Hidden Markov chains}~\cite{rabiner}.

In this section we derive a semi-analytical technique to calculate the KLD rate between hidden Markov chains. We focus on a simple case where the underlying Markov process is described by two observables $X$ and $Y$; however we only observe $X$ whose evolution is described by a hidden Markov chain. The KLD rate for the observable $X$ is 
\begin{equation}
\label{eq:hmcd}
d^{X}=\lim_{m\to\infty} \frac{1}{m} \sum_{x_1^m} p(x_1^m) \log
\frac {\sum_{y_1^m}p(x_1^m,y_1^m)}{\sum_{y_m^1}p(x_m^1,y_m^1)}.
\end{equation}
It is convenient to write $d^{X}$  as a difference between two terms,  $d^{X}=h_r^{X} -h^{X}$, where 
\begin{equation}
h^{X}  = -\lim_{m\to\infty} \frac{1}{m} \sum_{x_1^m}p(x_1^m)\log \sum_{y_1^m}p(x_1^m,y_1^m),
\end{equation}
 is called {\em Shannon entropy rate}, and
\begin{equation}
h_r^{X}  = -\lim_{m\to\infty} \frac{1}{m} \sum_{x_1^m}p(x_1^m)\log \sum_{y_m^1}p(x_m^1,y_m^1),
\end{equation}
{\em cross entropy rate}. Since the underlying process is Markovian, $p(x_1^m,y_1^m)$ factorizes and both Shannon and cross entropy can be expressed in terms of the trace of a product of random transition matrices $\mathbf{T}$~\cite{jacquet,holliday}. These are square $M\times M$ random matrices, where $M$ is the number of values that the variable $y$ can take on, and their entries  are given by
\begin{equation}
\mathbf{T}(x_1,x_2)_{y_1 y_2}= p(x_2, y_2 | x_1, y_1).
\label{transition_matrix}
\end{equation}
Note the different role played by each variable in this formalism: $x_{i}$  are parameters defining the matrix (making $\mathbf{T}$ a random matrix), whereas $y_{i}$ are subindices of the matrix elements. The Shannon and cross entropy can be expressed in terms of these matrices,

\begin{equation}
h^{X} = -\lim_{m\to \infty} \frac {1}{m} \left \langle \log \text{Tr}\left[ \prod_{i=1}^{m-1} \mathbf{T}(x_{i},x_{i+1}) \right]   \right \rangle,
\label{lf}
\end{equation}
\begin{equation}
h_r^{X} = - \lim_{m\to \infty} \frac {1}{m} \left \langle \log    \text{Tr}\left[\prod_{i=1}^{m-1}  \mathbf{T}(x_{m-i+1},x_{m-i})  \right ]  \right \rangle 
\label{lb}
\end{equation}
where $\langle \cdot\rangle$ denotes the average over  the random process $X_i$, which are weighted by $p(x_1^{m})$.  For sufficiently large $m$, Eqs.~(\ref{lf}) and (\ref{lb}) are self-averaging~\cite{holliday}, meaning that we do not need to calculate the average but just compute the trace for a single stationary trajectory. For any sufficiently long time series $\mathbf{x}=(\hat x_1,\hat x_2,\cdots,\hat x_n)$ with $n$ large, the following expressions converge to $-h$ and $-h_r$
almost surely, 

\begin{eqnarray}
\label{eq:lambda1}
\hat\lambda^{\mathbf{x}} =   \frac {1}{n}  \log \left \| \prod_{i=1}^{n-1} \mathbf{T}(\hat x_{i},\hat x_{i+1}) \right \|  \simeq  -h^{X}, \\
\label{eq:lambda2}
\hat\lambda^{\mathbf{\tilde x}} =\frac {1}{n}  \log \left   \|\prod_{i=1}^{n-1} \mathbf{T}(\hat x_{n-i+1},\hat x_{n-i}) \right \|  \simeq -h_r^{X}
\end{eqnarray}
where $\| \cdot \| $ is any matrix norm that satisfies $\|\mathbf{A}\cdot \mathbf{B} \| \leq \| \mathbf{A} \| \,\|\mathbf{B} \| $~\cite{holliday}.  
In particular, the trace satisfies this condition for positive matrices. In the context of random matrix theory, $\hat\lambda^{\mathbf{x}}$ and $\hat\lambda^{\mathbf{\tilde x}}$ are known as  \emph{maximum Lyapunov characteristic exponents}~\cite{prm} and measure the asymptotic rate of growth of a random vector when being multiplied by a random sequence of matrices.
In practice, we can estimate $d^{X}$ semi-analytically as
\begin{equation}
\label{eq:hmcfinal}
\hat{d}^{\mathbf{x}} = \hat \lambda^{\mathbf{x}}- \hat \lambda^{\mathbf{\tilde x}}.
\end{equation}
Here $\hat\lambda^{\mathbf{x}}$ and $\hat\lambda^{\mathbf{\tilde x}}$ are estimated using~\eqref{eq:lambda1} and \eqref{eq:lambda2} with a single time series $\mathbf{x}$ of size $n$, following a technique introduced in Ref.~\cite{prm}: we generate a random stationary time series $\mathbf{x}=\{\hat x_{1}^{n}\}$ and compute the matrices $\mathbf{T}$ analytically; then a random unitary vector is  multiplied by those matrices and normalized every $l$ data, keeping track of the normalization factor; finally the product of these factors divided by $n$ yields $\hat \lambda^{\mathbf{x}}$. For $\hat\lambda^{\mathbf{\tilde x}}$, the same procedure is repeated but using the reversed time series $\mathbf{\tilde x}=\{\hat x_{n}^{1}\}$. The technique is semi-analytical since the transition probabilities are known analytically but a single random stationary time series $\mathbf{x}$ is necessary to estimate $d^{X}$  with the multiplication of $n$ transition matrices that are chosen according to $\mathbf{x}$. 

Let us recall that the estimator $\hat{d}^{\mathbf{x}}$ cannot be applied to empirical time series unless we know the Markov model behind the data. Consequently, it is not useful in practical situations. However, we will use it to check the performance of the estimators introduced in the following section, which only need a single stationary time series to estimate the KLD and do not assume any knowledge of the dynamics generating these data. On the other hand, one can also get analytical approximations of Eqs.~(\ref{lf}) and~(\ref{lb}) by using the replica trick, in an analogous way as it has been done in Ref.~\cite{oliveira}. The calculation is cumbersome and is explained in Appendix \ref{sec:replica}. Both the semi-analytical and the replica calculations are used in Sec.~\ref{sec:flash} to check the accuracy of several empirical estimators of the KLD.

\section{Estimating KLD rates from single stationary trajectories}
\label{sec:estimators}

In previous sections, we calculated the KLD analytically (or semi-analytically) for series where we know in advance the dynamics of the underlying physical process.
We now investigate how the KLD rate can be estimated from a single empirical stationary trajectory,  obtained from a discrete stochastic
process whose dynamics is unknown. We call $\hat x_i$ the value of the $i-$th data of an empirical trajectory of $n$ data, which is denoted by $\mathbf{x}=\{\hat x_i\}_{i=1}^{n}$. There are two types of estimators in the literature:  \emph{plug-in} estimators, based on empirical counting of sequences of data, and estimators based on compression algorithms. In this section, we introduce a refinement of the these two methods and analyse their performance for a specific example in Sec. \ref{sec:flash}. 

\subsection{Plug-in estimators}
\label{sec:plug-in}

The simplest approach to estimate the KLD rate is known as the \emph{plug-in} method~\cite{wang}, which consists of an empirical estimation of the probabilities of sequences of $m$ data, $p(x_1^m)$, appearing in Eq.~(\ref{Dn}). The probability to observe the sequence $x_1^m$, $p(x_1^m)$, is estimated empirically from simply counting the number of times that $x_{1}^{m}$ appears in a single stationary trajectory $\mathbf{x}=(\hat x_1,\dots,\hat x_n)$ of size $n$. The empirical probability distribution is

\begin{equation}
\label{eq:plug-inp}
\hat{p}^{\mathbf{x}} (x_1^m)= \frac{1}{n-(m-1)} \sum_{p=1}^{n-(m-1)}  \delta_{\hat x_p,x_1} \cdots \delta_{\hat x_{p+(m-1)},x_m}
\end{equation}
Then an estimate of $D_m^X$ is obtained by plugging the empirical probability distribution into Eq.~(\ref{Dn}):

\begin{equation}
\label{eq:plug-inDn}
\hat{D}_m^{\mathbf{x}} = D[ \hat{p} ^{\mathbf{x}}(x_1^m) ||  \hat{p}^{\mathbf{x}}(x_m^1) ]  = \sum_{x_1,\cdots ,x_m} \hat{p}^{\mathbf{x}}(x_1^m) \log\frac{\hat{p}^{\mathbf{x}}(x_1^m)}{\hat{p}^{\mathbf{x}}(x_m^1)}.
\end{equation}
Note that  the probabilities in Eq.~\eqref{eq:plug-inDn} include the superscript $\mathbf{x}$ to emphasize that they are obtained empirically from a single stationary time series $\bf{x}$ and therefore depend on each particular realization. The simplest way   estimate $d^{X}$ would be by taking $\frac{\hat{D}_m^{\mathbf{x}}}{m}$ for $m$ as large as possible. However, this naive approach is not efficient. The empirical probability $\hat{p}^{\mathbf{x}}(x_1^m)$ ---and therefore $\hat{D}_m^{\mathbf{x}}$--- is less accurate as $m$ increases, because the number of possible substring $x_1^m$ increases exponentially and the statistics shortly becomes poor. It is convenient to find alternative expressions with a fast convergence. It turns out that the slope of $\hat D^{\mathbf{x}}_{m}$ as a function of $m$,
\begin{equation}
\label{eq:plug-indm} 
\hat d^{\mathbf{x}}_{m} = \hat D^{\mathbf{x}}_{m}-\hat D^{\mathbf{x}}_{m-1},
\end{equation}
also converges to the KLD rate but faster than $\frac{\hat{D}_m^{\mathbf{x}}}{m}$. Our plug-in estimator will be constructed as the limit
\begin{equation}
\hat d^{\mathbf{x}}=\lim_{m\to\infty}\hat d^{\mathbf{x}}_{m}.
\label{ddk}
\end{equation}
For a Markovian time series, as shown in Eq.~\eqref{d2}, the limit is reached for $m=2$, and using distributions of three or more data we only get redundant information: $\hat d^{\mathbf{x}}=\hat d^{\mathbf{x}}_{2}=\hat d^{\mathbf{x}}_{m}$, for any $m>2$. Therefore, $\hat d^{\mathbf{x}}=\hat d^{\mathbf{x}}_{2}$ is an excellent estimator of the KLD, $d^X$. If $\mathbf{x}$ is a $k$-th order Markov chain (i.e., it is Markovian when considering blocks of $k$ data $\{\hat x_1^k\}$), then the limit is reached for $m=k$, i.e., $\hat d^{\mathbf{x}}=\hat d^{\mathbf{x}}_{k}=\hat d^{\mathbf{x}}_{k+ 1}=\hat d^{\mathbf{x}}_{k+2}=\cdots$~\cite{rached}. The convergence of \eqref{ddk} is then expected to be fast if a time series can be approximated by a $k$-th order Markov chain.

If the trajectory $\mathbf{x}$ is sampled from a general non-Markovian process,  one needs further information to extrapolate $\hat d^{\mathbf{x}}_{m}$ for $m\to \infty$, specially when only  moderate values of $m$ can be reached. In the examples discussed below, we have found that convergence is well described by the following ansatz, proposed by Sch\"urmann and Grassberger~\cite{grass} to estimate Shannon entropy rate
\begin{equation}
\hat d^{\mathbf{x}}_{m} \simeq   \hat d^{\mathbf{x}}_{\infty} - c\frac{\log{m}}{m^{\gamma}}.
\label{ans}
\end{equation}
Here $c$ and $\gamma$ are parameters that, together with $\hat d^{\mathbf{x}}_{\infty}$, can be obtained  by fitting the empirical values of $\hat d^{\mathbf{x}}_m$ as a function of  $m$. The fitting parameter $\hat{d}^{\mathbf{x}}_\infty$ gives an estimation of the limit (\ref{ddk}).

This estimation method is efficient as long as there is sufficient statistics in the data, that is, if for every series $x_1^m$ that
occurs in the trajectory, its reverse $x_m^1$ is observed at least once. On the other hand, if  we find empirically $\hat{p}^\mathbf{x} (x_1^m) \neq 0 $
while $\hat p^{\mathbf{x}} (x_m^1)=0$ for at least one case, the argument of the logarithm in Eq.~(\ref{Dn}) diverges, yielding $\hat{d}_m^\mathbf{x}=\infty$. We can avoid this divergence by restricting the sum in $\hat{D}^\mathbf{x}_m$ to sequences $x_1^m$  whose reverse $x_m^1$  occur in the time series:
\begin{equation}
\hat D^{\mathbf{x}}_m\to \hat D_m^{\mathbf{x}\star} =\sum_{ (x_1^m)^* }  \hat{p}^{\mathbf{x}}(x_1^m) \log\frac{\hat{p}^{\mathbf{x}}(x_1^m)}{\hat{p}^{\mathbf{x}}(x_m^1)},
\label{nostat}
\end{equation}
where $(x_1^m)^* =  \{ x_1^m \;  |  \;  \hat p^{\mathbf{x}}(x_1^m)\neq 0 \; \text{and} \; \hat p^{\mathbf{x}}(x_m^1)\neq 0 \}$. 
With this restriction, a lower bound to $\hat D^{\mathbf{x}}_m$  is always obtained,  $ \hat D^{\mathbf{x}\star}_m < \hat D^{\mathbf{x}}_m $.

A different strategy is to artificially bias the empirical probabilities such that all of them become positive. Instead of the observed empirical frequencies,
we can use  the following biased frequencies~\cite{cai}
\begin{equation}
\hat{p}^{\mathbf{x}} (x_1^m) = \frac{n^{\mathbf{x}}(x_1^m)+\gamma}{\sum_{x_1^m} [n^{\mathbf{x}}(x_1^m)+\gamma]}.
\label{biased}
\end{equation}
Here $n^{\mathbf{x}} (x_1^m)$ is the number of observations of $x_1^m$ in $\mathbf{x}$ and $\gamma$ is the bias, which is a small number that prevents any of the probabilities to be zero, assigning a probability of order $\gamma/n$ to sequences that are not observed. The denominator in Eq.~\eqref{biased} ensures normalization of $\hat{p}^{\mathbf{x}} (x_1^m)$. 

% A similar
%result is obtained when estimating underlying probabilities using
%priors when there are few observations~\cite{jaynes}.
%

\subsection{Ziv-Merhav estimator}
\label{sec:ziv}

 Ziv and Merhav introduced in Ref.~\cite{ziv} an estimator of the KLD rate between two probability distributions based on compression algorithms. It consists on slicing or {\em parsing} stationary discrete time series into smaller parts according to a specific algorithm. The slicing produces a sequence of numbers (often called a {\em dictionary}) that contains the same data than the original series, but it is divided into subsequences, called {\em phrases}. The algorithms that are used are called {\em compression algorithms} because the number of phrases in which a time series $\mathbf{x}$ of $n$ numbers is parsed into is smaller than $n$. 
 
The estimator is defined in terms of two concepts which are now described, the compression length of a sequence and the cross parsing length between two different sequences.  Given a series $\mathbf{x}=x_{1}^{n}$,  its \emph{compression length} $c(x_{1}^{n})$ is defined as the number of distinct phrases in which it is parsed using the Lempel-Ziv (LZ) algorithm~\cite{lz78}. The LZ algorithm parses a series sequentially, such that each phrase that is added to the dictionary is the shortest distinct phrase that is not already in the dictionary. For example, let us consider the series $\mathbf{x}=x_1^{11}=(0, 1, 1, 1,1,0,0,0,1,1,0)$. The LZ sequential parsing for this example is as follows: First we store the first element of the sequence $x_1=0$ in the dictionary as it is empty, hence $\text{Dict}=\{0\}$. Then we read the next number, $x_2=1$, which is not already in the dictionary, so $x_2$ is added to the dictionary, $\text{Dict}=\{0| 1\}$. The next number in $x_1^{11}$ is $x_3=1$, which is already in the dictionary. Then we append to $x_3$ the next number of the sequence, $x_3^4=(1,1)$. This phrase is not in the dictionary and therefore it is parsed, $\text{Dict}=\{0 | 1 |(1,1)\}$.  By doing this for all the series $x_1^{11}$, we obtain the following dictionary of phrases  $\text{Dict}=\{0 |  1 |  (1,1) | (1,0) | (0,0) | (1,1,0) \}$. The compression length is the number of phrases that the dictionary contains once the series $\bf{x}$ is completely parsed, $c(x_1^{11})=6$ in this example. The compression length of a stationary time series is related to its Shannon entropy rate~\cite{cover} in the limit of infinitely long sequences:
\begin{equation}
\lim_{n\to\infty} \frac{c(x_1^n)\log c(x_1^n)}{n} = h^X.
\end{equation}
However, as $d^X=h_r^X-h^X$, we also require an estimator for $h_r^X$ in order to determine $d^{X}$. This is given in terms of another quantity called {\em cross parsing length}. The cross parsing of a series $x_1^n$ with respect to another sequence $z_1^n$ is obtained by parsing $x_1^n$ looking for the longest phrase that appears anywhere in $z_1^n$.  As an example, let us consider the cross parsing of $\mathbf{x}=x_1^{11}=(0, 1, 1, 1,1,0,0,0,1,1,0)$ with respect to another sequence $\mathbf{z}=z_{1}^{11}=(1,0,0,1,0,1,0,0,1,1,0)$. The first number in $\bf{x}$ is $x_1=0$, which is in $\bf{z}$. Therefore we append to $x_1$ the next number in $\bf{x}$, $x_1^2=(0, 1)$. This sequence is also somewhere in $\bf{z}$, more precisely it is equal to $z_3^4$ ,$z_5^6$ and $z_8^9$, so we append the next item in $\bf{x}$, $x_1^3=(0, 1,1)$. Again this sequence is somewhere in $\bf{z}$, $x_1^3=z_8^{10}$, and it is added to the dictionary, ~$\text{Dict}=\{(0, 1, 1)\}$ because $x_1^4$ is not equal to any subsequence of $z_1^{11}$. We repeat this procedure again starting from $x_4$ and the resulting dictionary is:  $\text{Dict}=\{ (0, 1 ,1) | (1, 1, 0) | (0, 0, 1, 1,  0)\} $. The cross parsing length is the number of parsed sequences, which in this example is equal to $c_r(x_1^{11} | z_1^{11})=3$. In Ref.~\cite{ziv} it is proved that the following quantity tends to the KLD rate between the probability distributions that generated the sequences $\mathbf{x}=x_1^n$ and $\mathbf{z}=z_1^n$, which we call $p^{X}$ and $q^{Z}$ respectively,

\begin{equation}\label{eq:lz1}
\lim_{n\to\infty} \frac{1}{n} [c_r(x_1^n|z_1^n)\log n - c(x_1^n)\log c(x_1^n)] = d(p^{X}||q^{Z}).
\end{equation}
We can estimate $d^X$ by using as inputs in the left-hand side of the above equation a stationary time series and its time reverse. The Ziv-Merhav estimator of $d^X$ when using a time series $\bf{x}$ of $n$ data is introduced as follows

\begin{equation}
\label{eq:lz2}
\hat d_{ZM}^{\mathbf{x}}  = \frac{1}{n} [c_r(x_1^n|x_n^1)\log n - c(x_1^n)\log c(x_1^n)],
\end{equation}
which converges to $d^X$ when $n\to\infty$, although the convergence is slow~\cite{ziv}.  This estimator has been used as a measure of distinguishability in several fields such as authorship attribution~\cite{cou} or biometric identification~\cite{coubio}. 

When the KLD rate between the probability distributions under consideration is small ($d^{X} \ll 1$), the estimation given by Eq.~\eqref{eq:lz1} can  be even negative~\cite{cou}. The estimator gives negative values in some cases because it mixes two types of parsing: the sequential parsing of the trajectory and the cross parsing, which is not sequential. We propose the following correction, which helps to solve this issue and improves the performance of the estimator. We first evaluate \eqref{eq:lz2} between different segments of the same trajectory. More precisely, we split $\bf{x}$ into two equal parts and apply the original estimator \eqref{eq:lz1}
\begin{equation}
\tilde d_{ZM}^{\mathbf{x}}  =  \frac{c_r(x_{n/2}^n|x_1^{n/2}) \log \frac{n}{2} - c(x_{n/2}^n)\log c(x_{n/2}^n)}{n/2}.
\end{equation}
If the time series is stationary, the two fragments, $x_1^{n/2}$ and $x_{n/2}^n$, are equivalent and $\tilde d_{ZM}^{\mathbf{x}} $ should vanish. However it is usually negative for finite $n$ and exhibits a slow convergence to zero for large $n$~\cite{cou}. Then, we define our estimator as
\begin{equation}\label{compestim}
\hat d^{\mathbf{x}}_c = \hat d_{ZM}^{\mathbf{x}} - \tilde d_{ZM}^{\mathbf{x}}   ,
\end{equation}
which still converges to $d$ when $n\to\infty$ and yields much better results for finite $n$, as we show with a simple example.

\begin{figure}
\includegraphics[width=4.5cm,angle=0]{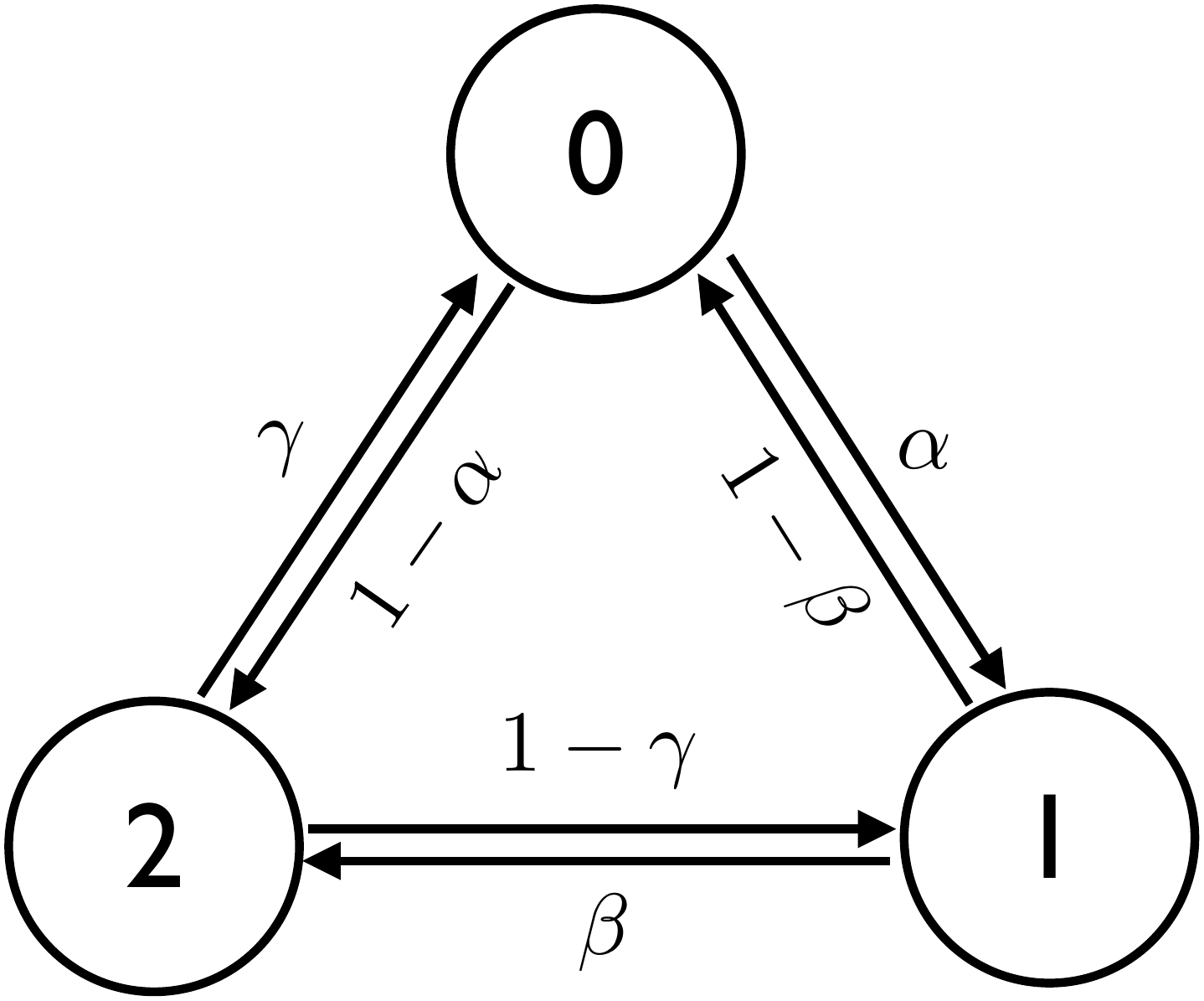}
\includegraphics[width=8cm]{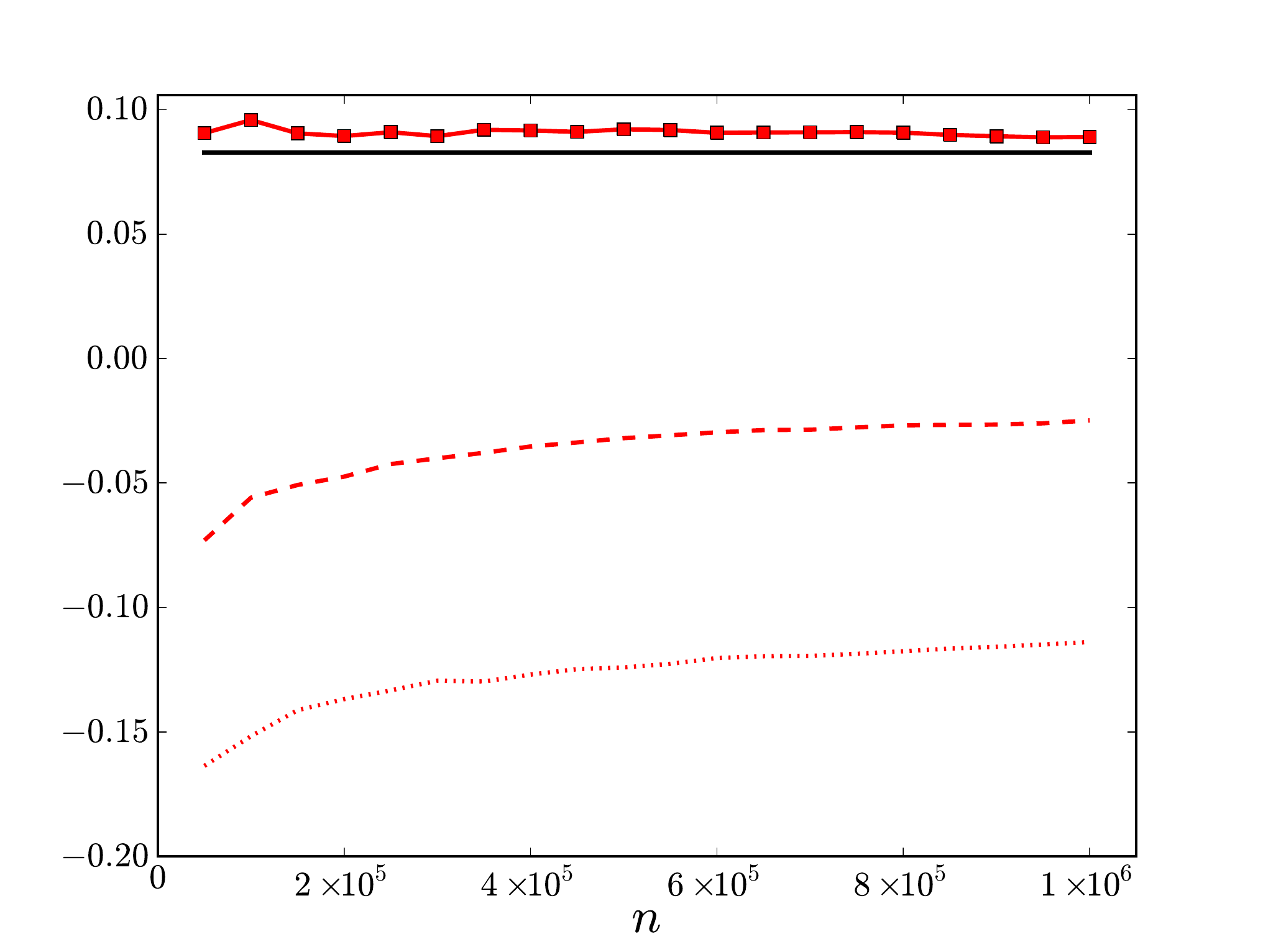}
\caption{ Sketch of the 3-state toy model used to check the accuracy of our compression estimator \eqref{compestim} and comparison between different compression estimators and the analytical value of $d^{X}$. The analytical value of $d^{X}$ for a model with $\alpha = 0.5, \beta=0.7 , \gamma = 0.6$  ($d^{X}=d^{X}_2=0.08278$) is indicated by the solid black line in the plot. We show the value of the compression estimators obtained from a single stationary time series $x_1^n$ as a function of the length  $n$: the Ziv-Merhav estimator $\hat d_{ZM}^{\mathbf{x}}$ (red dashed line), the bias $\tilde d_{ZM}^{\mathbf{x}}$ (red dotted line) and our estimator $\hat d_c^{\mathbf{x}} $ (red squares).}
 \label{3statefig}
\end{figure}

We perform a first validation of this estimator using the three-state model illustrated in Fig.~\ref{3statefig}.
Trajectories of the model are lists of numbers, $0$, $1$ or $2$,  representing  the three states of the system. The dynamics is Markovian with transition probabilities  given by $p_{0\to 1}=1-p_{1\to 0}=\alpha$, $p_{1\to 2}=1-p_{2\to 1}=\beta$ and $p_{2\to 0}=1-p_{0\to 2}=\gamma$. We call $X_i$ the stochastic process describing the state of the system and $\bf{x}$ a particular stationary time series,  e.g. $\mathbf{x}=( 0, 2, 1, 0, 1, 2, 1, 2, \cdots)$. This time series is reversible only when the three transition probabilities satisfy the Kolmogorov condition~\cite{kingman}, $\alpha\beta\gamma=(1-\alpha)(1-\beta)(1-\gamma)$. In Fig. \ref{3statefig} (lower plot)
 we compare the value of different compression estimators with the analytical value of $d^{X}$ as a function of the length of the empirical trajectory $n$. Since the trajectories described by  the state of the system are Markovian, $d^{X}$ only depends on transition probabilities: $d^{X}=d^{X}_2$.
 We see that the Ziv-Merhav estimator $\hat d_{ZM}^{\mathbf{x}}$ fails to estimate $d^{X}$ accurately when it is small ($d^{X}\simeq0.083$) and in some cases gives a negative value. The proposed estimator $\hat d^{\bf{x}}_c$, on the other hand, is significantly closer to the analytical result, although slightly overestimates its true value.

\section{Application: the discrete flashing ratchet}
\label{sec:flash}

\subsection{The model}

\begin{figure}
\includegraphics[width=8cm]{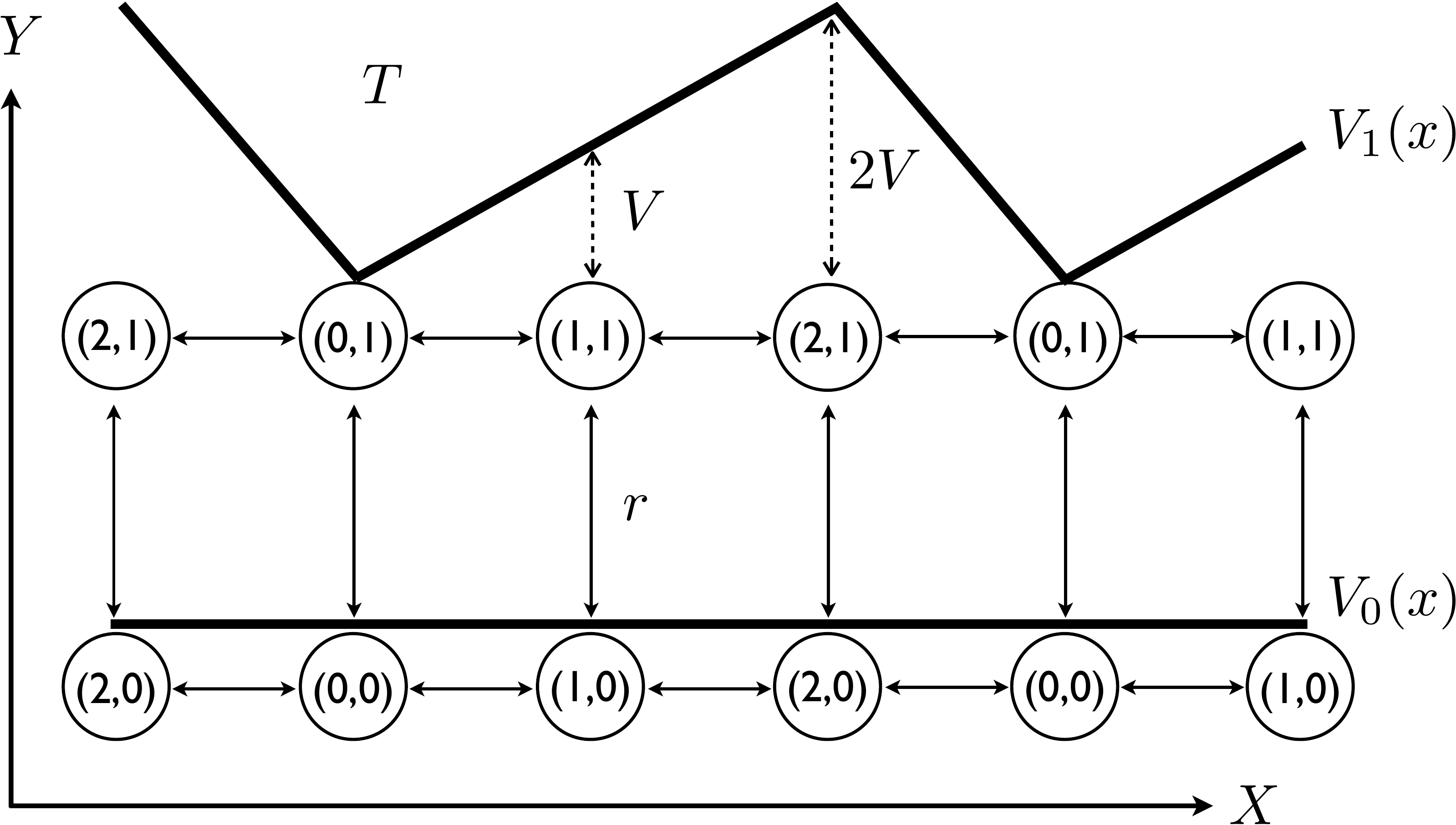}
\caption{Illustration of our discrete ratchet model. Particles are immersed in a thermal bath at temperature $T$ and move in one dimension in an asymmetric linear potential $V_1(x)$ of height $2V$ with periodic boundary conditions. The potential is switched on and off at a rate $r$, where $V_0(x)=0$ represents a flat potential, and the switching probability does not depend on the position of the particle. The state of the particle is represented by two random variables $(X,Y)$ indicated in the figure, where $X=\{0,1,2\}$ stands for the position of the particle whereas $Y=\{0,1\}$ for the state of the potential. Using this description, the system can be in six different states, $(0,0), (1,0), (2,0), (0,1), (1,1), (2,1)$.}
 \label{dr}
\end{figure}

We now apply the previous techniques to a specific example: a discrete flashing ratchet consisting of a Brownian particle moving on a one dimensional lattice~\cite{prost}. The particle is immersed in a thermal bath at temperature $T$ and moves in a periodic, linear, asymmetric potential of height $2V$, which is switched on and off at a constant rate $r$ (see Fig.~\ref{dr}).
Trajectories are denoted by two random observables: the position of the particle $X$ ($0$, $1$ or $2$) and the state of the potential $Y$ (ON, $Y=1$ or OFF, $Y=0$).

The particle evolves in continuous time according to a Master equation. The dynamics is described in terms of rates of spatial jumps and switching. For each possible transition except switches, i.e. $(x_1,y_1) \to (x_2,y_2)$ with $y_1=y_2=y$, we define a transition rate $k_{(x_1,y)\to (x_2,y)}$ obeying detailed balance,

\begin{equation}
k_{(x_1,y)\rightarrow (x_2,y)}=\exp\left[-\frac{V_y(x_2)-V_y(x_1)}{2kT}\right].
\end{equation}
When the potential is on ($y=1$), the value of the potential energy $V_1(x)$ is given in Fig.~\ref{dr}. When the potential is off, $V_0(x)=0$ for all $x$, and $k_{(x_1,0) \rightarrow (x_2,0)}=1$ for $ x_1\neq x_2 $. The switching rate does not depend on the position of the particle:  $k_{(x,y_1)\to (x,y_2)}=r$ for any value of $x$ and $y_1\neq y_2$, and consequently violates detailed balance, driving the system out of equilibrium.  

We simplify the analysis by   mapping  the dynamics onto a discrete-time process, a Markov chain.
To this end, we record in a time series $(\mathbf{x},\mathbf{y})=\{x_1^n,y_1^n\}$ just a list of the visited states, discarding any information about the time where jumps and switches occur. The resulting Markov chain is defined by the transition probabilities

 \begin{equation} 
 p[(x_2, y_2) |(x_1, y_1)] = \frac{ k_{(x_1, y_1) \to (x_2, y_2)}}{ \sum_{x_2 ,y_2}k_{(x_1, y_1)\to (x_2, y_2)}}.
\end{equation}
Since we discard any information about the transition times, we will focus along the rest of paper only on dissipation and KLD rates {\em per jump} or per data. 
For finite switching rate $r$, the ratchet rectifies the thermal fluctuations inducing a current to the left in Fig.~\ref{dr}~\cite{prost,ratchet}. The system obeys a local detailed balance condition, as described in Sec.~\ref{sec:markov}. The nonequilibrium nature of the switching can be interpreted in two alternative ways: one can imagine that it is activated by a thermal bath at infinite temperature or by an external agent~\cite{ratchet}. In either of the two interpretations, switching does not induce any entropy production (the bath needs an 
infinite amount of energy to change its entropy and the external agent does not produce any entropy change). Therefore, entropy is only produced when heat is dissipated to the bath at temperature $T$, which only occurs when the potential is on. The average entropy production (or dissipation) per data in the time series is then [cfr.~\eqref{markovequality}]

\begin{equation}
\label{eq:sdotfull}
\langle \dot{S} \rangle= \sum_{y=0,1}\; \sum_{x_{1},x_{2}=0,1,2}p[(x_{1},y);(x_{2},y)]\frac{ V_y(x_{1})-V_y(x_{2}) }{T},
\end{equation}
which is equal to the KLD rate when calculated for time series containing the information of both position and state of the system (which we call {\em full} information), $\langle \dot{S} \rangle=d^{X,Y}=d^{X,Y}_{2}$. We now analyze how  can $d$ be estimated using single stationary trajectories of this model, and how close is this estimation to the entropy production depending on the number of degrees of freedom of the system that are sampled in the time series.

\subsection{Full information}

\begin{figure}
\includegraphics[width=8cm]{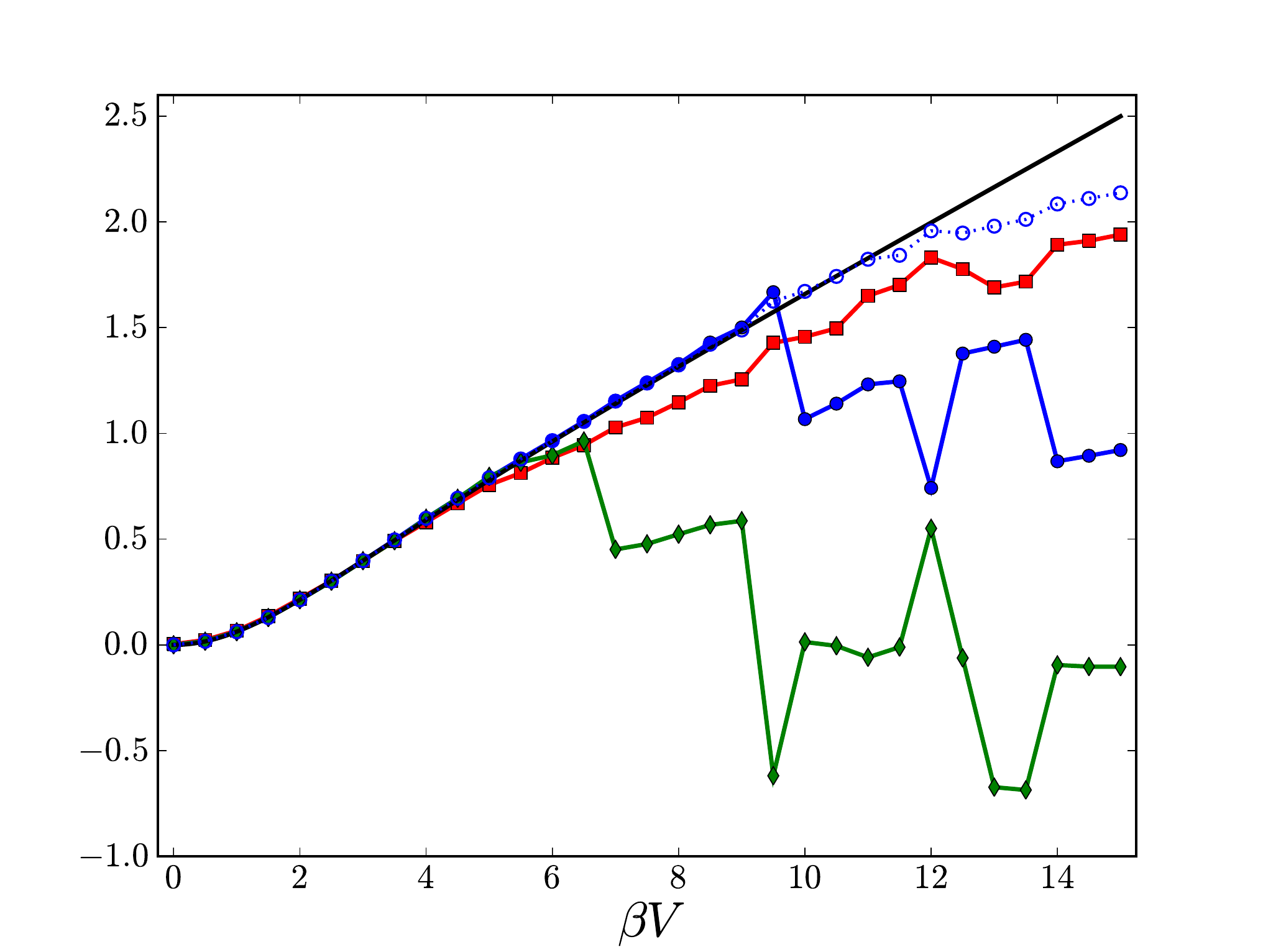}
\caption{Analytical value of the average dissipation per data in units of $kT$ (black line) as a function of  $\beta V$ in the flashing ratchet ($r=1$) and different estimators of $d^{X,Y}$. For each value of $\beta V$, estimators are obtained from a single stationary time series of $n=10^6$ data containing full information of the system (position, $X$, and state of the potential, $Y$): {\em Plug-in estimators}: $\hat d^{\bf{x,y}}_{2}$ (blue circles), $\hat d^{\bf{x,y}}_{3}$ (green diamonds), and  $\tilde d^{\bf{x,y}}_{2}$  using biased probabilities with $\gamma=1$ (blue open circles).  {\em Compression estimator}: $\hat d^{\bf{x,y}}_c$ (red squares). }
\label{dfull}
\end{figure}

Firstly, we investigate the estimation of the KLD rate when using full information of the system (the position of the particle $X$ and the state of the potential $Y$), and how close is this KLD rate to the actual entropy production of the process. In Fig.~\ref{dfull}
 we compare the actual dissipation and several empirical estimations of $d^{X,Y}$ for different values of the height of the potential, $V$. For each value of $V$ we simulate a single stationary time series of $n=10^{6}$ data that contains full information, and calculate the plug-in estimators $\hat d^{\bf{x,y}}_{2}$, $\hat d^{\bf{x,y}}_{3}$, as well as the compression-based estimator $\hat d^{\bf{x,y}}_{c}$.   

Since trajectories containing full information are Markovian, the plug-in estimator immediately converges to the dissipation $\hat d^{\bf{x,y}} _{2}=\hat d^{\bf{x,y}}= d^{X,Y}=\langle \dot{S}\rangle/k$ if there is enough statistics, which happens when $V$ is below or of order $kT$. If $V \gg kT$, the observation of the uphill jumps such as $(0,1)\to (1,1)$, $(0,1)\to (2,1)$, or $(1,1)\to(2,1)$ is very unlikely in a single stationary trajectory. A time series of $n$ data captures the statistics of jumps with probability well above $1/n$,  which amounts  to say energy jumps below $kT\log n$, ($kT\log 10^{6}\approx 14kT$ for the trajectory used in the figures).
 
 If, for instance, the transition $(0,1)\to (1,1)$ is missing in the trajectory, there is no way of estimating $p[(0,1);(1,1)]$ which contributes to two terms in $\hat d^{\bf{x,y}}_{2}$  [see Eq.~\eqref{Dn} for  $n=2$]. One of these two terms accounts for jumps $(0,1)\to (1,1)$,  which are very unlikely and their contribution to the total dissipation rate is negligible, and the other term accounts for jumps $(1,1)\to (0,1)$, whose probability is larger and therefore contribute more significantly to the entropy production. 

In Fig.~\ref{dfull}, $\hat d^{\bf{x,y}}_2$ (blue circles) and $\hat d^{\bf{x,y}}_3$ (green diamonds) have been calculated restricting the average to sequences (of two or three data respectively) whose reverse are also observed in the time series, as given by Eq.~\eqref{nostat}. The sudden drops in $\hat d^{\bf{x,y}}_{2}$ and $\hat d^{\bf{x,y}}_{3}$ are a consequence of lack of statistics in the trajectory. For the specific time series used in Fig.~\ref{dfull}, the lack of statistics starts at $\beta V\simeq 10$ for $\hat d^{\bf{x,y}}_{2}$ and arises earlier for $\hat d^{\bf{x,y}}_3$ because the three-data sampling space is bigger and it is easier that some transitions $(x_1,y_1) \to (x_2,y_2) \to (x_3,y_3)$ do not appear while their reverse do.

A more efficient way of dealing with the missing sequences is incorporating a small bias to the empirical probabilities, as described in Eq.~(\ref{biased}). This is equivalent to assigning a probability of order $1/n$ to those transitions that are not observed in a time series of $n$ data. Figure~\ref{dfull} shows $\tilde d^{\bf{x,y}}_2$  with a bias $\gamma=1$ (blue open circles), which is able to extend the accuracy of the estimation even when there is lack of statistics.

Although in the case of Markovian series with a finite number of states the most convenient strategy is to use the plug-in estimator, we include for comparison the compression estimator $\hat d^{\bf{x,y}}_c$ (red squares) which gives accurate values of the dissipation for weak potentials. Furthermore, the compression estimator is  better than some plug-in estimators even for strong potentials, since it does not exhibit sudden jumps due to lack of statistics.

\subsection{Partial information}

We now analyze the performance of our estimators when there is not access to the full description of the system. As in~\cite{roldan}, we assume that only the position of the ratchet $X$ is observable. Accordingly, we simulate trajectories containing full information, and we remove the information of the state afterwards, $(\mathbf{x,y})\to\mathbf{x}$. The resulting time series $\mathbf{x}=\{x_1^n\}$ is not Markovian and hence  the limit \eqref{ddk} is not reached for small values of $m$. In this case, we proceed by obtaining  $\hat d ^{\mathbf{x}}_m$ for $m$ as large as possible and fit the resulting values to the ansatz \eqref{ans}.

\begin{figure}
\includegraphics[width=8cm]{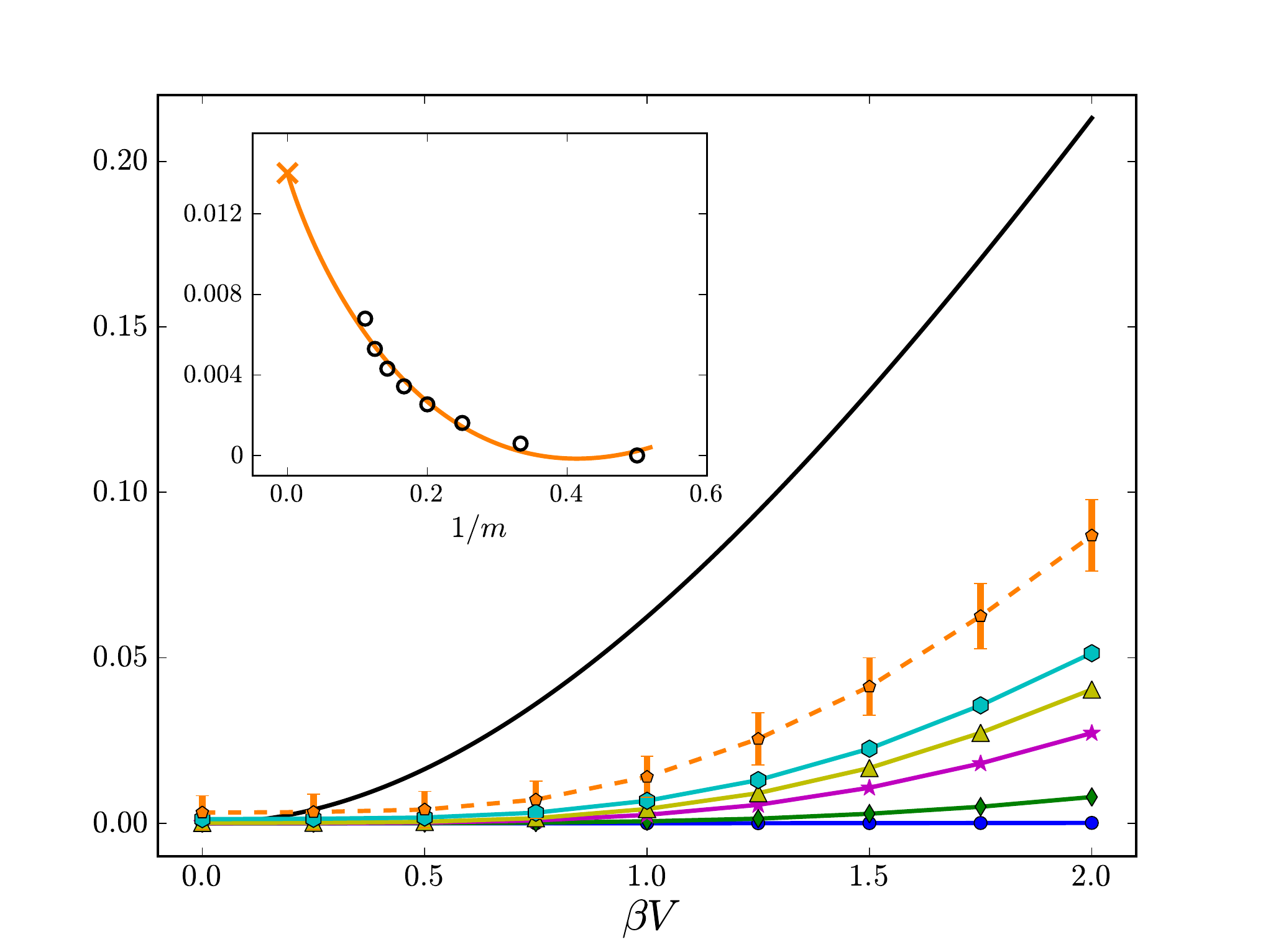}
\caption{Average dissipation per data (black line) and plug-in estimators of $d^{X}$ using partial information given by the position ($X$) for a discrete flashing ratchet with $r=1$. For each value of $\beta V$, we calculate estimators from a single stationary time series of $n=10^{7}$ data containing partial information: $\hat d^{\bf{x}}_2$(blue circles), $\hat d^{\bf{x}}_3$ (green diamonds), $\hat d^{\bf{x}}_5$ (purple stars), $\hat d^{\bf{x}}_7$ (yellow triangles), $\hat d^{\bf{x}}_9$ (cyan hexagons) and the result from the fit $\hat d^{\bf{x}}_\infty$ (orange pentagons with error bars and connected by a dashed line). \emph{Inset}: $\hat d^{\bf{x}}_m$ as a function of $1/m$ for $m=1,\cdots, 9$ for $\beta V=1$ (open black circles) and the fit to the ansatz (orange line). The $y-$intercept of the fit is indicated by an orange cross and it is equal to $\hat d^{\bf{x}}_\infty$.}\label{ansfig}
\end{figure}

\begin{figure}
\includegraphics[width=8cm]{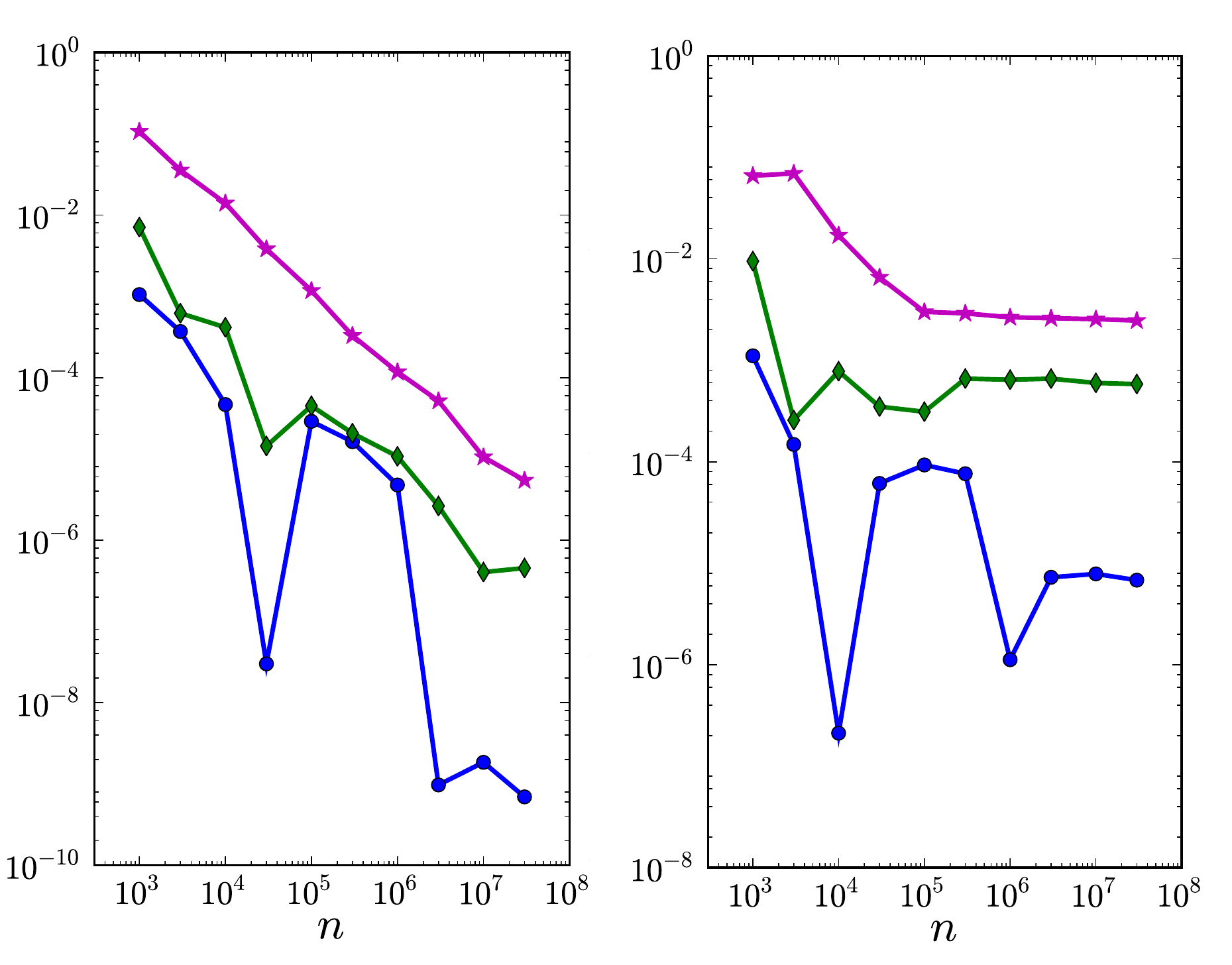}
\caption{Scaling of plug-in estimators of $d^{X}$, $\hat d^{\mathbf{x}}_m$, with the size of the time series $n$, for a flashing ratchet ($r=1$), for  $\beta V=0$ (left) and  $\beta V=1$ (right): $\hat d^{\bf{x}}_2$ (blue circles), $\hat d^{\bf{x}}_3$ (green diamonds) and $\hat d^{\bf{x}}_5$ (purple stars).  We simulate a single stationary trajectory $\bf{x}$ of $10^7$ data and calculate the estimators for subsequences containing the first $n$ data of $\bf{x}$ .}
\label{scaling}
\end{figure}

 We have generated trajectories of size $n=10^7$ for values of $V$ that range from $0$ to $2kT$. Once we remove the information of the state of the potential from these time series,  we are able to estimate $\hat d^{\bf{x}}_{m}$ up to $m=9$ with no lack of  statistics. Figure~\ref{ansfig}
shows the plug-in estimators $\hat d^{\bf{x}}_m$ for $m=2,3,5,7,9$ and the extrapolation $\hat d^{\bf{x}}_{\infty}$ (orange pentagons connected by a dashed line to guide the eye) resulting from the fit to the ansatz~(\ref{ans}). For each value of $\beta V$, we fit $\hat d^{\bf{x}}_{m}$ as a function of $m$ for $m=2,3,\cdots , 9$ to Eq.~(\ref{ans}) using the curve fitting tool available in \text{MATLAB}, which provides a robust least-squares fit with bisquare weights as described in~\cite{matlab}. The fit itself for a particular value of the potential, $\beta V=1$, is shown in the inset of Fig.~\ref{ansfig}.
  Our ansatz reproduces the dependence of $\hat d^{\bf{x}}_m$ with $m$ but the final estimator $\hat d^{\bf{x}}_{\infty}$ still bounds significantly from below the actual dissipation (black solid line in Fig.~\ref{ansfig}). Nevertheless, plug-in estimators clearly distinguish between equilibrium and NESS, even with partial information. In equilibrium ($V=0$), the trajectories are reversible and all the estimators vanish, $\hat d^{\bf{x}}_m=0$ for $m=2,\cdots , 9$,  whereas for the NESS ($V>0$) they detect the irreversibility of  the process yielding $\hat d^{\bf{x}}_m > 0 $ for all $m$. This is illustrated in Fig.~\ref{scaling},
 where we plot the dependence of the plug-in estimators with the size of the trajectory. For  $\beta V=0$, $\hat d^{\bf{x}}_2, \hat d^{\bf{x}}_3$ and $\hat d^{\bf{x}}_5$ tend to zero when increasing the number of data whereas they saturate to a positive value in the NESS ($\beta V=1$).

There are two possible origins for the discrepancy between $\hat d^{\bf{x}}_{\infty}$ and the dissipation: either (i) our fit underestimates  the actual KLD rate $d^{X}$ of the trajectory; or (ii) the bound \eqref{main_equation} is not tight. To address this question we need to calculate the actual value of $d^{X}$. Since the position of the ratchet $\bf{x}$ is a hidden Markov chain, we can calculate its KLD rate $d^{X}$  semi-analytically, using the Lyapunov exponents (\ref{eq:lambda1},\ref{eq:lambda2}) introduced in Sec.~\ref{sec:hmc}.

\begin{figure}
\includegraphics[width=8cm]{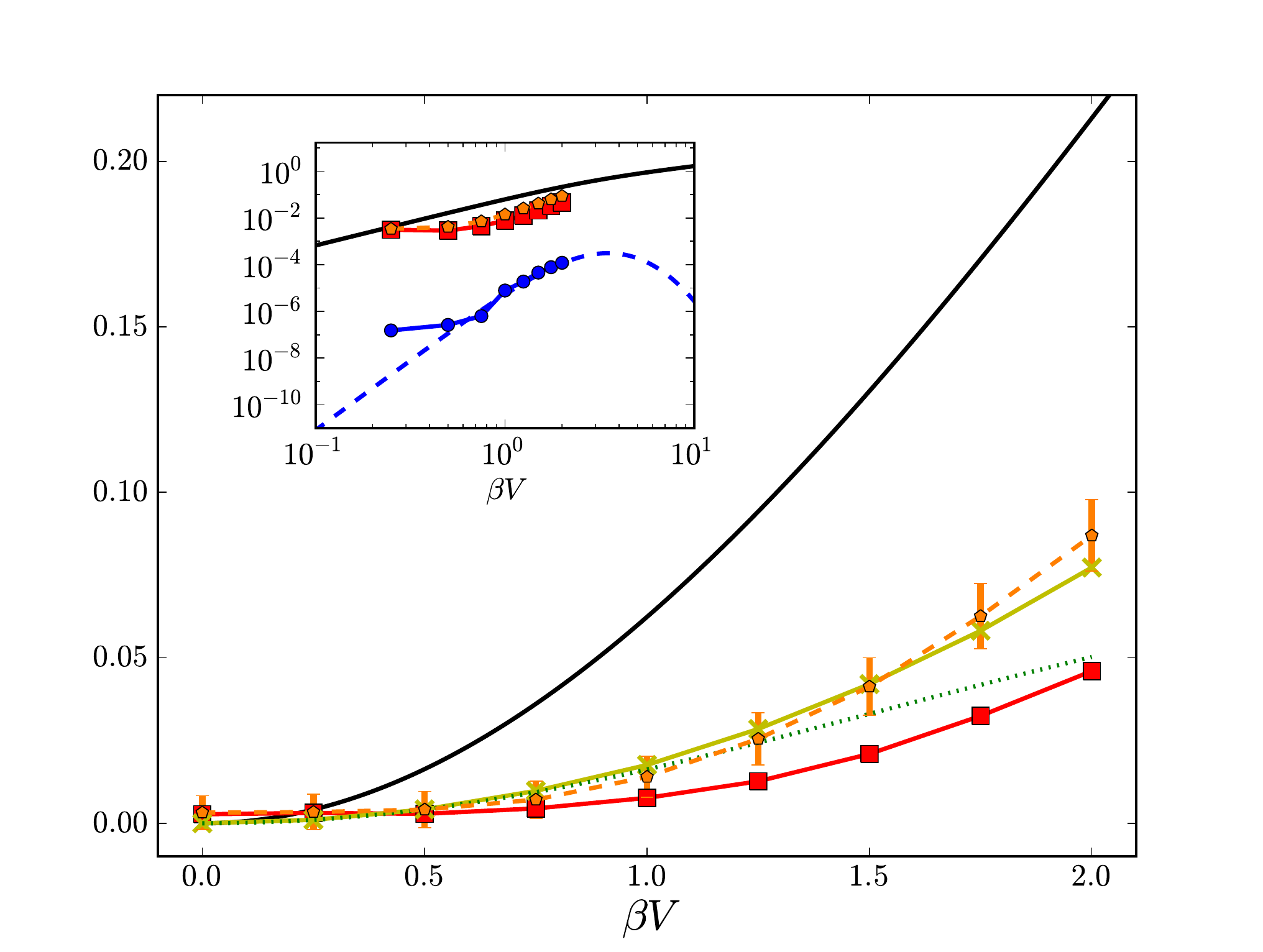}
\caption{Average dissipation per data (black line) and different estimators of $d^{X}$ for a flashing ratchet described with partial information  ($r=1$, $n=10^7$ data)
as a function of $\beta V$:  $\hat d^{\bf{x}}_\infty$ (orange dashed pentagons) $\hat d^{\bf{x}}_c$ (red squares), replica estimation of $d^{X}$  (green dotted line)
 and semi-analytical value of $d^{X}$  (yellow crosses). \emph{Inset}: Dependence of the average dissipation (black line), $\hat d^{\bf{x}}_2$ (analytical values in blue dashed line), $\hat d^{\bf{x}}_c$ and $\hat d^{\bf{x}}_\infty$ on $\beta V$ in the vicinity of $\beta V=0$.}
 \label{estimators}
\end{figure}

In Fig.~\ref{estimators} 
we show the value of the semi-analytical calculation of $d^{X}$ using the norm of transition matrices, Eq.~\eqref{eq:hmcfinal}, which is not significantly different to the empirical estimation $\hat d^{\bf{x}}_{\infty}$. We therefore conclude that $\hat d^{\bf{x}}_{\infty}$ is a good estimation of $d^{X}$, but still $d^{X}$ only yields a lower bound to dissipation whose accuracy is in principle hard to determine. This is an expected result, since the position of a particle in a flashing ratchet does not obey the Gallavotti-Cohen theorem~\cite{lacoste}. 

Summarizing, although $\hat d^{\bf{x}}_{\infty}$ turns out to be a good estimator of $d^{X}$, using only information of the position we only get a lower bound to the dissipation. We also show in Fig.~\ref{estimators} the value of $\hat d^{\bf{x}}_c$, which is well below the plug-in estimator $\hat d^{\bf{x}}_{\infty}$. The compression estimator $\hat d^{\bf{x}}_{c}$  lies between $\hat d^{\bf{x}}_7$ and $\hat d^{\bf{x}}_9$ (not shown in the plot), indicating that it is only able to capture correlations up to size 8. For completeness, we include the
calculation of $d^{X}$ based on the replica trick (see appendix \ref{sec:replica}). It yields a tight bound for $V<kT$, but departs from $d^{X}$ for larger values of $V$. This deviation is caused by the estimation of the limits in Eqs.~(\ref{hrep},\ref{hrrep}),  where we take ${\alpha\to 0}$ when $\alpha$ is defined only for integer values, one of the standard drawbacks of the replica trick~\cite{prm}.

Although our estimators give low values of the dissipation when using partial information, they still capture the asymptotic behavior for $V$ small. Entropy production decreases  as
$V^{2}$ when $V\to 0$, so do plug-in estimators $\hat d^{\bf{x}}_3,\cdots , \hat d^{\bf{x}}_9$, $\hat d^{\bf{x}}_{\infty}$, and the compression estimator $\hat d^{\bf{x}}_c$. Some of them are plotted in the inset of Fig.~\ref{estimators} (inset). On the other hand, $\hat d^{\bf{x}}_{2}\propto V^{6}$, since the current is $J\propto V^{3}$ in this case [see Eq.
\eqref{D2J}]. Recall that  calculating $\hat d^{\bf{x}}_{2}$ is equivalent to estimating the entropy production using currents and standard linear irreversible thermodynamics, as shown in Eq.~\eqref{D2J}. It is then remarkable that the estimators involving the statistics of three or more data are able to reproduce qualitatively the behavior of the dissipation in cases where linear thermodynamics fails.

\begin{figure}
\includegraphics[width=8cm]{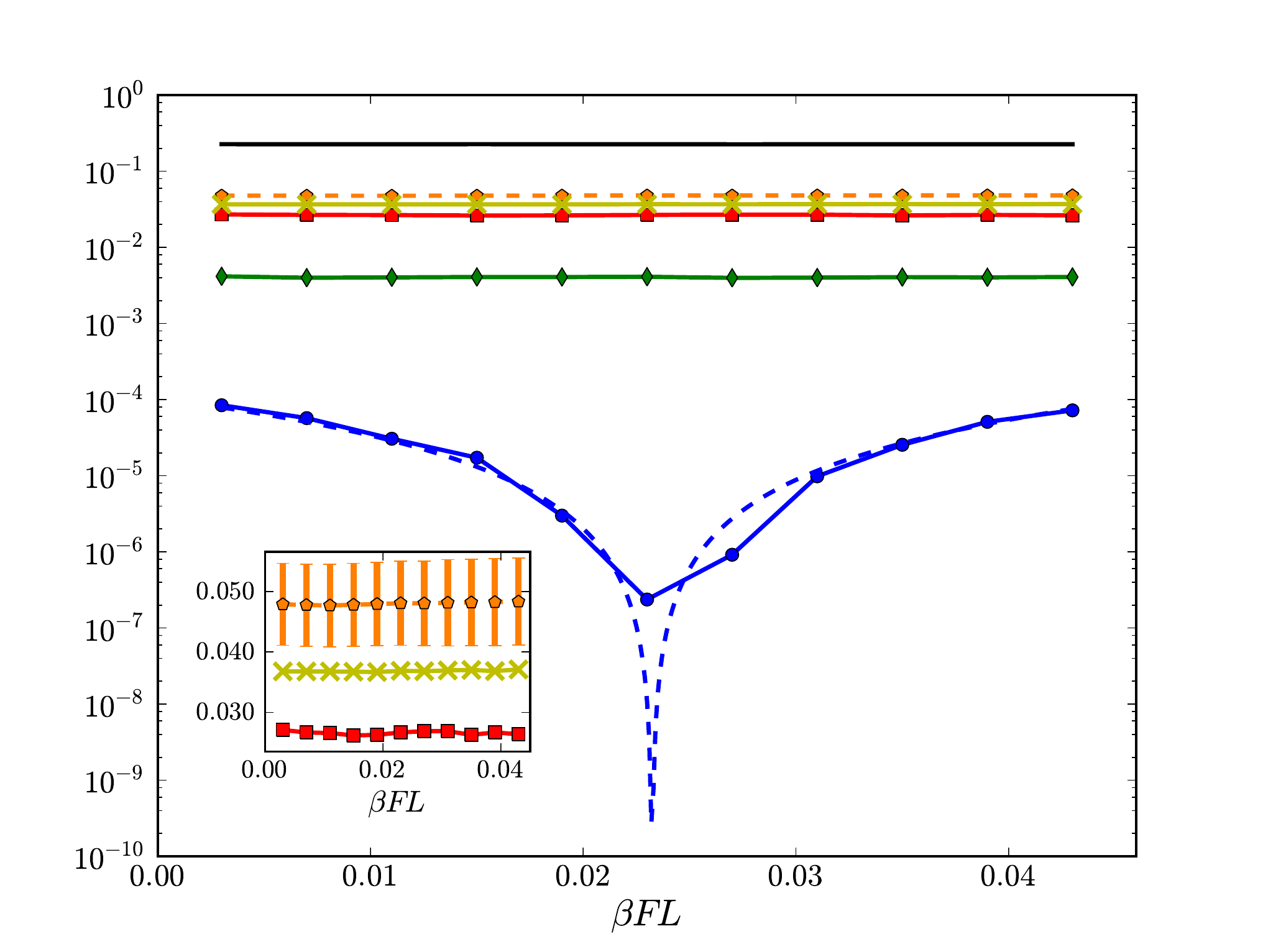}
\caption{Average dissipation per data (in units of $kT$) in the flashing ratchet (with $r=2$, and $\beta V=2$) and different estimations of $d^{X}$ obtained from a single time series of $n=10^{7}$ data containing partial information (position) as a function of the external force $F$: analytical value of the average dissipation (black line), $\hat d^{\bf{x}}_{2}$ (blue circles, analytical values in blue dashed line), $\hat d^{\bf{x}}_{3}$ (green diamonds), $\hat d^{\bf{x}}_c$ (red squares), semi-analytical calculation of $d^{X}$ (yellow crosses) and $\hat d^{\bf{x}}_\infty$ (orange hexagons). The minimum in $\hat d^{\bf{x}}_2$ corresponds to the stalling force. \emph{Inset}: $\hat d^{\bf{x}}_c$, semi-analytical value of $d^{X}$  and $\hat d^{\bf{x}}_\infty$ as a function of the external force.}
\label{stalling}
\end{figure}

The improvement observed when using  the plug-in estimators of higher order than $\hat d^{\bf{x}}_{2}$ is more dramatic in a NESS which does not exhibit observable currents in $X$. In this case $\hat d^{\bf{x}}_{2}=0$ but using higher order statistics we can still detect the time irreversibility of the trajectory~\cite{roldan}. This happens for example if we add to the flashing ratchet an external force $F$ opposite to the current, i.e., pointing in the positive $x-$direction. The force modifies the energy landscape and consequently the spatial transition rates $k_{(x_1 y) \to (x_2,y)}$ by a factor $\exp[\beta FL_{(x_1,y);(x_2,y)}/2]$, $L_{(x_1,y);(x_2,y)}$ being the spatial distance that separates the two points $(x_1,y)$ and $(x_2,y)$. Here $L_{(x_1,y);(x_2,y)}$ is defined positive if the jump $(x_1,y)\to (x_2,y)$ points in the same direction as the force (i.e. to the right), and negative otherwise. At the \emph{stalling force} $F_{\rm stall}$, the current is canceled by the force and the system does not move on average when it is described only by $X$, but still dissipates energy. If we only have access to the information of the position, the system looks like it is in equilibrium: the spatial current vanishes, and so does $\hat d^{\bf{x}}_2$, as shown in Fig.~\ref{stalling}.
 However, there is a finite dissipation (black line in the figure) and the corresponding irreversibility is captured by the statistics of substrings of length $3$ or more. Although $d^{X}$ is below the real dissipation by an order of magnitude (see the semi-analytical value of $d^X$, yellow crosses in Fig.~\ref{stalling})  , it does not exhibit any sensible change at stall force. Finally, both $\hat d^{\bf{x}}_\infty$ and  $\hat d^{\bf{x}}_c$ provide estimates of $d^{X}$ which are correct within one order of magnitude (see the inset of Fig. \ref{stalling}).

\section{Conclusions}
\label{sec:concl}

We have shown that it is possible to estimate the entropy production rate by analyzing statistical properties of a time series observed in a NESS. The Kullback-Leibler divergence
(KLD) per data between the time series and its time reversed is a lower bound to the entropy production rate.

We have introduced two estimators of this KLD rate, one based on empirical frequencies and another on compression algorithms, and we have checked their performance in
a specific example: a discrete flashing ratchet. We show that the KLD is a powerful tool to identify nonequilibrium states and to estimate the entropy production of a process, if this entropy production is of order of the Boltzmann constant. We have also shown that the bound given by the KLD can detect a non-zero dissipation even when the data does not
exhibit any measurable flows.

Let us summarize our results by presenting a ``recipe'' to estimate the KLD from an experimental time series recorded from a discrete system in a NESS. If the number of possible states of the system is small enough, the best approach is to calculate the plug-in estimators $\hat d^{\bf{x}}_m$~\eqref{eq:plug-indm} and then check the convergence when $m$ increases. The possible lack of statistics can be circumvented using a small artificial bias, as discussed in Sec.~\ref{sec:plug-in}. If $\hat d^{\bf{x}}_m$ saturates for some value $m^*$, then the time series is an $m^*$-th order Markov process and $\hat d^{\bf{x}}=\hat d^{\bf{x}}_{m^{*}}$. Otherwise, we can use the ansatz \eqref{ans} and obtain $\hat d^{\bf{x}}_\infty$ which is a good estimate of the KLD rate.

A second and complementary approach is the use of the compression estimator introduced in Sec.~\ref{sec:ziv}. The estimator yields correct results in the examples that we have analyzed, but there is no clue about the corresponding error. Nevertheless, the compression estimator could be the only possible approach if the number of
states of the time series is large. In this case, the calculation of empirical probability distributions $\hat p(x_1^m)$ would be unfeasible even for short substrings.

Another possible strategy for systems with many states (or described by real-valued observables) is to consider time asymmetric functionals of the data, which reduce the number of observables, and hence the number of states, but keep information about the irreversibility of the series. In any case, the estimation of KLD and the extension of our results to processes described by continuous data is an open problem, which will be relevant in many practical situations, especially to analyze data coming from biological systems.

Finally, let us mention that, as in the case of Landauer's principle, the KLD could also be used to ascertain the minimal entropy production associated with a specific behavior, such as spatiotemporal patterns, excitable systems, etc. This in turn may influence the design of optimal devices with functionalities given by these behaviors.

\begin{acknowledgments}
We acknowledge fruitful discussions with J. M. Horowitz and financial support from Grant MOSAICO  (Spanish Government) and MODELICO (Comunidad Aut\'onoma de Madrid, Spain). ER acknowledges the funding from {\em Becas de la Caixa para estudios de M\'aster en Espa\~na} and Beca FPU (Spanish Government).
\end{acknowledgments}

\appendix

\section{Calculation of the KLD rate for hidden Markov chains using replica trick}
\label{sec:replica}

The semi-analytical calculation of the KLD rate for a specific case of hidden Markov chains was discussed in Sec.~\ref{sec:hmc}. We now introduce another technique to calculate Eq.~\eqref{eq:hmcd} using a mathematical technique called {\em replica trick}. To this end, we first consider the expression of $d^{X}$ in terms of Shannon and cross entropy rates, $d^X=h_r^X-h^X$. We define the matrix resulting from the multiplication of $m$ transition matrices [defined in Eq.~(\ref{transition_matrix})] chosen according to $x_1^m$ by
\begin{equation}
\label{eq:transprod}
\mathbf{T}(x_1^m) =  \prod_{i=1}^{m-1} \mathbf{T}(x_{i},x_{i+1}) .
\end{equation}
Shannon entropy rate $h^{X}$ can be rewritten by susbtituing~\eqref{eq:transprod} into Eq.~(\ref{lf}),
\begin{equation}
\label{eq:hap}
h^{X} = -\lim_{m\to \infty} \frac {1}{m} \left \langle \log \text{Tr} \mathbf{T}(x_1^m) \right \rangle.
\end{equation}
%\begin{equation}
%h_r^{X} = -\lim_{n\to \infty} \frac {1}{n} \left \langle \log \text{Tr} \mathbf{T}(x_n^1) \right \rangle.
%\end{equation}
The analytical calculation of the average $\left \langle \log \text{Tr} \mathbf{T}(x_1^m) \right \rangle$ is cumbersome and it can only be done semi-analytically, as we explained in Sec.~\ref{sec:hmc}. However, we can express this average in terms of $\left \langle \text{Tr} \mathbf{T}(x_1^m) \right \rangle$, which can be calculated analytically. The mathematical technique to do this is called replica trick and it was introduced to calculate free energies in spin glasses~\cite{Hemmen}.  For our specific example, the trick is given by the following expression:

\begin{equation}
\langle \log \text{Tr} \mathbf{T}(x_1^m)  \rangle = \lim_{\alpha \to 0} \frac{d}{d\alpha}\log \langle [ \text{Tr} \mathbf{T}(x_1^m)  ] ^{\alpha} \rangle.
\label{replicadef}
\end{equation}
 Reference~\cite{oliveira} shows how to apply this technique when $\mathbf{T}(x_1^m)$ is equal to a product of random matrices which are chosen following a Markovian process. In our case, an underlying Markovian process defined by two random variables, $X$ and $Y$, defines the order of the matrices that are multiplied in $\mathbf{T}(x_1^m)$. We now apply the technique described in~\cite{oliveira} to calculate $h^{X}$. If we define the {\em generalized Lyapunov exponent} of degree $\alpha$~\cite{prm} as
 
\begin{equation}
L^{X}_{\alpha}  = \lim_{m\to\infty} \frac{1}{m} \log \left \langle [\text{Tr}  \mathbf{T}(x_1^m) ] ^{\alpha}  \right\rangle,
\label{gen_lyap}
\end{equation}
and we take into account replica trick~(\ref{replicadef}), Shannon entropy rate~\eqref{eq:hap} is given by

\begin{equation}
\label{eq:hxrep}
h^{X}=-\lim_{\alpha \to 0} \frac{d}{d\alpha} L^{X}_{\alpha}.
\end{equation}
Now we consider the following property: Given a matrix $\mathbf{A}$ and a positive integer $\alpha$, $(\text{Tr}\mathbf{A})^\alpha =\text{Tr}(\mathbf{A} ^{\otimes \alpha} ) $,  where $\mathbf{A}^{\otimes \alpha} =\underbrace{\mathbf{A}\otimes \mathbf{A}\otimes\cdots \otimes \mathbf{A}}_{\alpha \; \text{times}}$. Using this property, the average in Eq.~(\ref{gen_lyap}) reads

\begin{equation}
\label{eq:trprop}
 \left \langle  [\text{Tr}  \mathbf{T}(x_1^m) ] ^{\alpha}  \right\rangle = \left \langle  \text{Tr} [ \mathbf{T}(x_1^m)^{\otimes \alpha} ]  \right\rangle=  \text{Tr} \left \langle  \mathbf{T}(x_1^m)^{\otimes \alpha}  \right \rangle. 
\end{equation}
Since the tensor power of a  product of matrices factorizes, $(\mathbf{A}\mathbf{B}\mathbf{C})^{\otimes
\alpha} = \mathbf{A}^{\otimes \alpha}  \mathbf{B}^{\otimes \alpha} \mathbf{C}^{\otimes \alpha} $, Eq.~\eqref{eq:trprop} can be rewritten,

\begin{equation}
\label{eq:trprop2}
\left \langle  [\text{Tr}  \mathbf{T}(x_1^m) ] ^{\alpha}  \right\rangle = \text{Tr} \sum_{x_1^m,y_1^m} \prod_{i=1}^{m-1} \mathbf{T}(x_i,x_{i+1})_{y_i,y_{i+1}} \mathbf{T}(x_i,x_{i+1})^{\otimes \alpha} .
\end{equation}
We now define a block matrix $\mathcal{T}(\alpha)$, where each block is a transition matrix $\mathbf{T} (x_1,x_2)^{\otimes \alpha + 1}$. The matrix elements of $\mathcal{T}(\alpha)$ are therefore:

\begin{equation}
\label{eq:blockmat}
\mathcal{T}(\alpha)_{x_1,y_1,x_2,y_2} = [\mathbf{T} (x_1,x_2) ^{\otimes \alpha + 1}]_{y_1,y_2}.
\end{equation}
Using~\eqref{eq:trprop2} and~\eqref{eq:blockmat} in~(\ref{gen_lyap}), we see that $L^{X}_\alpha$ is dominated by the  largest eigenvalue of $\mathcal{T}(\alpha)$ which we call $\tau(\alpha)$,
\begin{equation}
L^{X}_\alpha = \lim_{m\to\infty} \frac{1}{m}\log \text{Tr}[\mathcal{T}(\alpha)^{m-1}] = \log  \tau (\alpha),
\end{equation}
yielding,
\begin{equation}
h^{X} = -\lim_{\alpha \to 0} \frac{d}{d\alpha} \log  \tau (\alpha).
\label{hrep}
\end{equation}
The above limit cannot be calculated analytically because the tensor powers in $\mathcal{T} (\alpha)$ are only defined for integer values of $\alpha$. Therefore we approximate the limit $\alpha \to 0$ by an estimation of the slope of  $L_\alpha^{X}$ as a function of $\alpha$ close to $\alpha = 0$, which is given by~\cite{prm}

\begin{equation}
\label{eq:hestrep}
\hat h^{X}= 2L^{X}_1-\frac{L^{X}_2}{2}= 2 \log  \tau (1)-\frac{ \log  \tau (2)}{2}.
\end{equation}
We obtain an equivalent result for $h_r^{X}$ by replacing $\mathbf{T}(x_1^m)$  in Eq.~\eqref{eq:hap} by the product of transition matrices but ordered according to the time-reversed series $x_m^1$, $\mathbf{T}(x_m^1)$. Defining the following matrix 
\begin{equation}
\mathcal{T}_r (\alpha)_{x_1,y_1,x_2,y_2}
= [\mathbf{T} (x_2,x_1)^{T} \otimes \mathbf{T} (x_1,x_2) ^{\otimes \alpha}]_{y_1,y_2},
\end{equation}
and being  $\tau_r(\alpha)$ the largest eigenvalue of $\mathcal{T}_r (\alpha)$, we get
\begin{equation}
h_r^{X}  = -\lim_{\alpha \to 0} \frac{d}{d\alpha} \log  \tau _r (\alpha).
\label{hrrep}
\end{equation}
In practice,  we also need to approximate the limit $\alpha\to 0$ in the above equation using Eq.~\eqref{eq:hestrep} but replacing $\tau$ by $\tau_r$,
\begin{equation}
\label{eq:hrestrep}
\hat h_r^{X}=  2 \log  \tau_r (1)-\frac{ \log  \tau_r (2)}{2}.
\end{equation}
Finally, the estimation of $d^{X}$ for this kind of series using replica trick, which is shown in Fig.~\ref{estimators} (green dotted line), is obtained with the difference between Eqs.~\eqref{eq:hrestrep} and~\eqref{eq:hestrep},
\begin{equation}
\hat d^{X}=\hat h^{X}_r- \hat h^{X}=2 \log  \frac{\tau_r (1)}{\tau (1)}+\frac{1}{2} \log \frac{\tau (2)}{ \tau_r (2)}. 
\end{equation}
\bibliography{refs28jul.bib}

\begin{thebibliography}{48}
\expandafter\ifx\csname natexlab\endcsname\relax\def\natexlab#1{#1}\fi
\expandafter\ifx\csname bibnamefont\endcsname\relax
  \def\bibnamefont#1{#1}\fi
\expandafter\ifx\csname bibfnamefont\endcsname\relax
  \def\bibfnamefont#1{#1}\fi
\expandafter\ifx\csname citenamefont\endcsname\relax
  \def\citenamefont#1{#1}\fi
\expandafter\ifx\csname url\endcsname\relax
  \def\url#1{\texttt{#1}}\fi
\expandafter\ifx\csname urlprefix\endcsname\relax\def\urlprefix{URL }\fi
\providecommand{\bibinfo}[2]{#2}
\providecommand{\eprint}[2][]{\url{#2}}

\bibitem[{\citenamefont{Kawai et~al.}(2007)\citenamefont{Kawai, Parrondo, and
  den Broeck}}]{kpb}
\bibinfo{author}{\bibfnamefont{R.}~\bibnamefont{Kawai}},
  \bibinfo{author}{\bibfnamefont{J.~M.~R.} \bibnamefont{Parrondo}},
  \bibnamefont{and} \bibinfo{author}{\bibfnamefont{C.~V.} \bibnamefont{den
  Broeck}}, \bibinfo{journal}{Phys. Rev. Lett.} \textbf{\bibinfo{volume}{98}},
  \bibinfo{eid}{080602} (\bibinfo{year}{2007}).

\bibitem[{\citenamefont{Parrondo et~al.}(2009)\citenamefont{Parrondo, den
  Broeck, and Kawai}}]{njp}
\bibinfo{author}{\bibfnamefont{J.~M.~R.} \bibnamefont{Parrondo}},
  \bibinfo{author}{\bibfnamefont{C.~V.} \bibnamefont{den Broeck}},
  \bibnamefont{and} \bibinfo{author}{\bibfnamefont{R.}~\bibnamefont{Kawai}},
  \bibinfo{journal}{New J. Phys.} \textbf{\bibinfo{volume}{11}},
  \bibinfo{pages}{073008} (\bibinfo{year}{2009}).

\bibitem[{\citenamefont{Cover and Thomas}(2006)}]{cover}
\bibinfo{author}{\bibfnamefont{T.~M.} \bibnamefont{Cover}} \bibnamefont{and}
  \bibinfo{author}{\bibfnamefont{J.~A.} \bibnamefont{Thomas}},
  \emph{\bibinfo{title}{Elements of information theory}}
  (\bibinfo{publisher}{Wiley}, \bibinfo{address}{Hoboken, New Jersey},
  \bibinfo{year}{2006}), \bibinfo{edition}{2nd} ed.

\bibitem[{\citenamefont{Crooks and Sivak}(2011)}]{crooks2011}
\bibinfo{author}{\bibfnamefont{G.~E.} \bibnamefont{Crooks}} \bibnamefont{and}
  \bibinfo{author}{\bibfnamefont{D.~A.} \bibnamefont{Sivak}},
  \bibinfo{journal}{Journal of Statistical Mechanics: Theory and Experiment}
  \textbf{\bibinfo{volume}{2011}}, \bibinfo{pages}{P06003}
  (\bibinfo{year}{2011}).

\bibitem[{\citenamefont{Maes and Netocny}(2003)}]{maes}
\bibinfo{author}{\bibfnamefont{C.}~\bibnamefont{Maes}} \bibnamefont{and}
  \bibinfo{author}{\bibfnamefont{K.}~\bibnamefont{Netocny}},
  \bibinfo{journal}{Journal of Statistical Physics}
  \textbf{\bibinfo{volume}{110}}, \bibinfo{pages}{269} (\bibinfo{year}{2003}).

\bibitem[{\citenamefont{Jarzynski}(2006)}]{jarz2006}
\bibinfo{author}{\bibfnamefont{C.}~\bibnamefont{Jarzynski}},
  \bibinfo{journal}{Phys. Rev. E} \textbf{\bibinfo{volume}{73}},
  \bibinfo{pages}{046105} (\bibinfo{year}{2006}).

\bibitem[{\citenamefont{Gaspard}(2004)}]{gaspard_markov}
\bibinfo{author}{\bibfnamefont{P.}~\bibnamefont{Gaspard}},
  \bibinfo{journal}{Journal of Statistical Physics}
  \textbf{\bibinfo{volume}{117}}, \bibinfo{pages}{599} (\bibinfo{year}{2004}).

\bibitem[{\citenamefont{Zolfaghari et~al.}(2010)\citenamefont{Zolfaghari, Zare,
  and Mirza}}]{zolfaghari}
\bibinfo{author}{\bibfnamefont{P.}~\bibnamefont{Zolfaghari}},
  \bibinfo{author}{\bibfnamefont{S.}~\bibnamefont{Zare}}, \bibnamefont{and}
  \bibinfo{author}{\bibfnamefont{B.}~\bibnamefont{Mirza}},
  \bibinfo{journal}{Phys. Rev. E} \textbf{\bibinfo{volume}{82}},
  \bibinfo{pages}{052104} (\bibinfo{year}{2010}).

\bibitem[{\citenamefont{Rold\'an and Parrondo}(2010)}]{roldan}
\bibinfo{author}{\bibfnamefont{E.}~\bibnamefont{Rold\'an}} \bibnamefont{and}
  \bibinfo{author}{\bibfnamefont{J.~M.~R.} \bibnamefont{Parrondo}},
  \bibinfo{journal}{Phys. Rev. Lett.} \textbf{\bibinfo{volume}{105}},
  \bibinfo{pages}{150607} (\bibinfo{year}{2010}).

\bibitem[{\citenamefont{Lebowitz and Spohn}(1999)}]{lebowitz}
\bibinfo{author}{\bibfnamefont{J.~L.} \bibnamefont{Lebowitz}} \bibnamefont{and}
  \bibinfo{author}{\bibfnamefont{H.}~\bibnamefont{Spohn}},
  \bibinfo{journal}{Journal of Statistical Physics}
  \textbf{\bibinfo{volume}{95}}, \bibinfo{pages}{333} (\bibinfo{year}{1999}),
  ISSN \bibinfo{issn}{0022-4715}.

\bibitem[{\citenamefont{{Mazonka} and {Jarzynski}}(1999)}]{mazonka}
\bibinfo{author}{\bibfnamefont{O.}~\bibnamefont{{Mazonka}}} \bibnamefont{and}
  \bibinfo{author}{\bibfnamefont{C.}~\bibnamefont{{Jarzynski}}}
  (\bibinfo{year}{1999}), \eprint{arXiv:cond-mat/9912121}.

\bibitem[{\citenamefont{van Zon et~al.}(2004)\citenamefont{van Zon, Ciliberto,
  and Cohen}}]{vanZon}
\bibinfo{author}{\bibfnamefont{R.}~\bibnamefont{van Zon}},
  \bibinfo{author}{\bibfnamefont{S.}~\bibnamefont{Ciliberto}},
  \bibnamefont{and} \bibinfo{author}{\bibfnamefont{E.~G.~D.}
  \bibnamefont{Cohen}}, \bibinfo{journal}{Phys. Rev. Lett.}
  \textbf{\bibinfo{volume}{92}}, \bibinfo{pages}{130601}
  (\bibinfo{year}{2004}).

\bibitem[{\citenamefont{Seifert}(2005)}]{seifert}
\bibinfo{author}{\bibfnamefont{U.}~\bibnamefont{Seifert}},
  \bibinfo{journal}{Phys. Rev. Lett.} \textbf{\bibinfo{volume}{95}},
  \bibinfo{pages}{040602} (\bibinfo{year}{2005}).

\bibitem[{\citenamefont{Andrieux et~al.}(2007)\citenamefont{Andrieux, Gaspard,
  Ciliberto, Garnier, Joubaud, and Petrosyan}}]{ciliberto}
\bibinfo{author}{\bibfnamefont{D.}~\bibnamefont{Andrieux}},
  \bibinfo{author}{\bibfnamefont{P.}~\bibnamefont{Gaspard}},
  \bibinfo{author}{\bibfnamefont{S.}~\bibnamefont{Ciliberto}},
  \bibinfo{author}{\bibfnamefont{N.}~\bibnamefont{Garnier}},
  \bibinfo{author}{\bibfnamefont{S.}~\bibnamefont{Joubaud}}, \bibnamefont{and}
  \bibinfo{author}{\bibfnamefont{A.}~\bibnamefont{Petrosyan}},
  \bibinfo{journal}{Phys. Rev. Lett.} \textbf{\bibinfo{volume}{98}},
  \bibinfo{pages}{150601} (\bibinfo{year}{2007}).

\bibitem[{\citenamefont{Horowitz and Jarzynski}(2009)}]{jordan}
\bibinfo{author}{\bibfnamefont{J.}~\bibnamefont{Horowitz}} \bibnamefont{and}
  \bibinfo{author}{\bibfnamefont{C.}~\bibnamefont{Jarzynski}},
  \bibinfo{journal}{Phys. Rev. E} \textbf{\bibinfo{volume}{79}},
  \bibinfo{pages}{021106} (\bibinfo{year}{2009}).

\bibitem[{\citenamefont{Gomez-Marin
  et~al.}(2008{\natexlab{a}})\citenamefont{Gomez-Marin, Parrondo, and den
  Broeck}}]{footprints}
\bibinfo{author}{\bibfnamefont{A.}~\bibnamefont{Gomez-Marin}},
  \bibinfo{author}{\bibfnamefont{J.~M.~R.} \bibnamefont{Parrondo}},
  \bibnamefont{and} \bibinfo{author}{\bibfnamefont{C.~V.} \bibnamefont{den
  Broeck}}, \bibinfo{journal}{EPL} \textbf{\bibinfo{volume}{82}},
  \bibinfo{pages}{50002} (\bibinfo{year}{2008}{\natexlab{a}}).

\bibitem[{\citenamefont{Kurchan}(1998)}]{kurchan}
\bibinfo{author}{\bibfnamefont{J.}~\bibnamefont{Kurchan}},
  \bibinfo{journal}{Journal of Physics A} \textbf{\bibinfo{volume}{31}},
  \bibinfo{pages}{3719} (\bibinfo{year}{1998}).

\bibitem[{\citenamefont{Garnier and Ciliberto}(2005)}]{garnier}
\bibinfo{author}{\bibfnamefont{N.}~\bibnamefont{Garnier}} \bibnamefont{and}
  \bibinfo{author}{\bibfnamefont{S.}~\bibnamefont{Ciliberto}},
  \bibinfo{journal}{Phys. Rev. E} \textbf{\bibinfo{volume}{71}},
  \bibinfo{pages}{060101} (\bibinfo{year}{2005}).

\bibitem[{\citenamefont{Martin et~al.}(2001)\citenamefont{Martin, Hudspeth, and
  J\"ulicher}}]{julicher}
\bibinfo{author}{\bibfnamefont{P.}~\bibnamefont{Martin}},
  \bibinfo{author}{\bibfnamefont{A.~J.} \bibnamefont{Hudspeth}},
  \bibnamefont{and}
  \bibinfo{author}{\bibfnamefont{F.}~\bibnamefont{J\"ulicher}},
  \bibinfo{journal}{P. Natl. Acad. Sci. USA} \textbf{\bibinfo{volume}{98}},
  \bibinfo{pages}{14380} (\bibinfo{year}{2001}).

\bibitem[{\citenamefont{Amann et~al.}(2010)\citenamefont{Amann, Schmiedl, and
  Seifert}}]{seifert2}
\bibinfo{author}{\bibfnamefont{C.~P.} \bibnamefont{Amann}},
  \bibinfo{author}{\bibfnamefont{T.}~\bibnamefont{Schmiedl}}, \bibnamefont{and}
  \bibinfo{author}{\bibfnamefont{U.}~\bibnamefont{Seifert}},
  \bibinfo{journal}{J. Chem. Phys.} \textbf{\bibinfo{volume}{132}},
  \bibinfo{pages}{041102} (\bibinfo{year}{2010}).

\bibitem[{\citenamefont{Kennel}(2004)}]{kennel}
\bibinfo{author}{\bibfnamefont{M.~B.} \bibnamefont{Kennel}},
  \bibinfo{journal}{Phys. Rev. E} \textbf{\bibinfo{volume}{69}},
  \bibinfo{pages}{056208} (\bibinfo{year}{2004}).

\bibitem[{\citenamefont{Coutinho and Figueiredo}(2005)}]{cou}
\bibinfo{author}{\bibfnamefont{D.~P.} \bibnamefont{Coutinho}} \bibnamefont{and}
  \bibinfo{author}{\bibfnamefont{M.~A.} \bibnamefont{Figueiredo}},
  \emph{\bibinfo{title}{Pattern Recognition and Image Analysis}}, vol.
  \bibinfo{volume}{3523} of \emph{\bibinfo{series}{Lecture Notes in Computer
  Science}} (\bibinfo{publisher}{Springer Berlin / Heidelberg},
  \bibinfo{year}{2005}).

\bibitem[{\citenamefont{Rached et~al.}(2004)\citenamefont{Rached, Alajaji, and
  Campbell}}]{rached}
\bibinfo{author}{\bibfnamefont{Z.}~\bibnamefont{Rached}},
  \bibinfo{author}{\bibfnamefont{F.}~\bibnamefont{Alajaji}}, \bibnamefont{and}
  \bibinfo{author}{\bibfnamefont{L.~L.} \bibnamefont{Campbell}},
  \bibinfo{journal}{IEEE T. Inform. Theory} \textbf{\bibinfo{volume}{50}},
  \bibinfo{pages}{917} (\bibinfo{year}{2004}).

\bibitem[{\citenamefont{Wang et~al.}(2005)\citenamefont{Wang, Kulkarni, and
  Verdu}}]{wang}
\bibinfo{author}{\bibfnamefont{Q.}~\bibnamefont{Wang}},
  \bibinfo{author}{\bibfnamefont{S.}~\bibnamefont{Kulkarni}}, \bibnamefont{and}
  \bibinfo{author}{\bibfnamefont{S.}~\bibnamefont{Verdu}},
  \bibinfo{journal}{IEEE Transactions on Information Theory}
  \textbf{\bibinfo{volume}{51}}, \bibinfo{pages}{3064 } (\bibinfo{year}{2005}).

\bibitem[{\citenamefont{Budka et~al.}(2011)\citenamefont{Budka, Gabrys, and
  Musial}}]{budka}
\bibinfo{author}{\bibfnamefont{M.}~\bibnamefont{Budka}},
  \bibinfo{author}{\bibfnamefont{B.}~\bibnamefont{Gabrys}}, \bibnamefont{and}
  \bibinfo{author}{\bibfnamefont{K.}~\bibnamefont{Musial}},
  \bibinfo{journal}{Entropy} \textbf{\bibinfo{volume}{13}},
  \bibinfo{pages}{1229 } (\bibinfo{year}{2011}).

\bibitem[{\citenamefont{Ziv and Merhav}(1993)}]{ziv}
\bibinfo{author}{\bibfnamefont{J.}~\bibnamefont{Ziv}} \bibnamefont{and}
  \bibinfo{author}{\bibfnamefont{N.}~\bibnamefont{Merhav}},
  \bibinfo{journal}{IEEE T. Inform. Theory} \textbf{\bibinfo{volume}{39}}
  (\bibinfo{year}{1993}).

\bibitem[{\citenamefont{Crooks}(1999)}]{crooks}
\bibinfo{author}{\bibfnamefont{G.~E.} \bibnamefont{Crooks}},
  \bibinfo{journal}{Phys. Rev. E} \textbf{\bibinfo{volume}{60}},
  \bibinfo{pages}{2721} (\bibinfo{year}{1999}).

\bibitem[{\citenamefont{Maes}(2003)}]{maes2}
\bibinfo{author}{\bibfnamefont{C.}~\bibnamefont{Maes}}, \bibinfo{journal}{Sem.
  Poincar\'e} \textbf{\bibinfo{volume}{2}} (\bibinfo{year}{2003}).

\bibitem[{\citenamefont{Cohen and Gallavotti}(1999)}]{cohen}
\bibinfo{author}{\bibfnamefont{E.}~\bibnamefont{Cohen}} \bibnamefont{and}
  \bibinfo{author}{\bibfnamefont{G.}~\bibnamefont{Gallavotti}},
  \bibinfo{journal}{J. Stat. Phys.} \textbf{\bibinfo{volume}{96}},
  \bibinfo{pages}{1343} (\bibinfo{year}{1999}).

\bibitem[{\citenamefont{Cleuren et~al.}(2008)\citenamefont{Cleuren, Willaert,
  Engel, and Van~den Broeck}}]{koen}
\bibinfo{author}{\bibfnamefont{B.}~\bibnamefont{Cleuren}},
  \bibinfo{author}{\bibfnamefont{K.}~\bibnamefont{Willaert}},
  \bibinfo{author}{\bibfnamefont{A.}~\bibnamefont{Engel}}, \bibnamefont{and}
  \bibinfo{author}{\bibfnamefont{C.}~\bibnamefont{Van~den Broeck}},
  \bibinfo{journal}{Phys. Rev. E} \textbf{\bibinfo{volume}{77}},
  \bibinfo{pages}{022103} (\bibinfo{year}{2008}).

\bibitem[{\citenamefont{Gomez-Marin
  et~al.}(2008{\natexlab{b}})\citenamefont{Gomez-Marin, Parrondo, and Van~den
  Broeck}}]{alexpre}
\bibinfo{author}{\bibfnamefont{A.}~\bibnamefont{Gomez-Marin}},
  \bibinfo{author}{\bibfnamefont{J.~M.~R.} \bibnamefont{Parrondo}},
  \bibnamefont{and} \bibinfo{author}{\bibfnamefont{C.}~\bibnamefont{Van~den
  Broeck}}, \bibinfo{journal}{Phys. Rev. E} \textbf{\bibinfo{volume}{78}},
  \bibinfo{pages}{011107} (\bibinfo{year}{2008}{\natexlab{b}}).

\bibitem[{\citenamefont{Landauer}(2000)}]{landauer}
\bibinfo{author}{\bibfnamefont{R.}~\bibnamefont{Landauer}},
  \bibinfo{journal}{IBM J. Res. Dev.} \textbf{\bibinfo{volume}{44}},
  \bibinfo{pages}{261} (\bibinfo{year}{2000}).

\bibitem[{\citenamefont{Andrieux and Gaspard}(2008)}]{gasp}
\bibinfo{author}{\bibfnamefont{D.}~\bibnamefont{Andrieux}} \bibnamefont{and}
  \bibinfo{author}{\bibfnamefont{P.}~\bibnamefont{Gaspard}},
  \bibinfo{journal}{P. Natl. Acad. Sci. USA} \textbf{\bibinfo{volume}{105}},
  \bibinfo{pages}{9516} (\bibinfo{year}{2008}).

\bibitem[{\citenamefont{{Parrondo} and {de Cisneros}}(2002)}]{ratchet}
\bibinfo{author}{\bibfnamefont{J.~M.~R.} \bibnamefont{{Parrondo}}}
  \bibnamefont{and} \bibinfo{author}{\bibfnamefont{B.~J.} \bibnamefont{{de
  Cisneros}}}, \bibinfo{journal}{App. Phys. A} \textbf{\bibinfo{volume}{75}},
  \bibinfo{pages}{179} (\bibinfo{year}{2002}).

\bibitem[{\citenamefont{Rabiner and Juang}(1986)}]{rabiner}
\bibinfo{author}{\bibfnamefont{L.}~\bibnamefont{Rabiner}} \bibnamefont{and}
  \bibinfo{author}{\bibfnamefont{B.}~\bibnamefont{Juang}},
  \bibinfo{journal}{ASSP Magazine, IEEE} \textbf{\bibinfo{volume}{3}},
  \bibinfo{pages}{4 } (\bibinfo{year}{1986}).

\bibitem[{\citenamefont{Jacquet et~al.}(2008)\citenamefont{Jacquet, Seroussi,
  and Szpankowski}}]{jacquet}
\bibinfo{author}{\bibfnamefont{P.}~\bibnamefont{Jacquet}},
  \bibinfo{author}{\bibfnamefont{G.}~\bibnamefont{Seroussi}}, \bibnamefont{and}
  \bibinfo{author}{\bibfnamefont{W.}~\bibnamefont{Szpankowski}},
  \bibinfo{journal}{Theor. Comput. Sci.} \textbf{\bibinfo{volume}{395}},
  \bibinfo{pages}{203} (\bibinfo{year}{2008}).

\bibitem[{\citenamefont{T.~Holliday and Goldsmith}(2004)}]{holliday}
\bibinfo{author}{\bibfnamefont{P.~G.} \bibnamefont{T.~Holliday}}
  \bibnamefont{and}
  \bibinfo{author}{\bibfnamefont{A.}~\bibnamefont{Goldsmith}},
  \bibinfo{journal}{Submitted to IEEE Trans. Inform. Theory}
  (\bibinfo{year}{2004}).

\bibitem[{\citenamefont{Crisanti et~al.}(1993)\citenamefont{Crisanti, Paladin,
  and Vulpiani}}]{prm}
\bibinfo{author}{\bibfnamefont{A.}~\bibnamefont{Crisanti}},
  \bibinfo{author}{\bibfnamefont{G.}~\bibnamefont{Paladin}}, \bibnamefont{and}
  \bibinfo{author}{\bibfnamefont{A.}~\bibnamefont{Vulpiani}},
  \emph{\bibinfo{title}{Products of Random Matrices in Statistical Physics}}
  (\bibinfo{publisher}{Springer Series in Solid State Sciences},
  \bibinfo{year}{1993}), ISBN \bibinfo{isbn}{0387565752}.

\bibitem[{\citenamefont{de~Oliveira and Petri}(1996)}]{oliveira}
\bibinfo{author}{\bibfnamefont{M.~J.} \bibnamefont{de~Oliveira}}
  \bibnamefont{and} \bibinfo{author}{\bibfnamefont{A.}~\bibnamefont{Petri}},
  \bibinfo{journal}{Phys. Rev. E} \textbf{\bibinfo{volume}{53}},
  \bibinfo{pages}{2960} (\bibinfo{year}{1996}).

\bibitem[{\citenamefont{Schurmann and Grassberger}(1996)}]{grass}
\bibinfo{author}{\bibfnamefont{T.}~\bibnamefont{Schurmann}} \bibnamefont{and}
  \bibinfo{author}{\bibfnamefont{P.}~\bibnamefont{Grassberger}},
  \bibinfo{journal}{Chaos} \textbf{\bibinfo{volume}{6}}, \bibinfo{pages}{414}
  (\bibinfo{year}{1996}).

\bibitem[{\citenamefont{Cai et~al.}(2006)\citenamefont{Cai, Kulkarni, and
  Verdu}}]{cai}
\bibinfo{author}{\bibfnamefont{H.}~\bibnamefont{Cai}},
  \bibinfo{author}{\bibfnamefont{S.}~\bibnamefont{Kulkarni}}, \bibnamefont{and}
  \bibinfo{author}{\bibfnamefont{S.}~\bibnamefont{Verdu}},
  \bibinfo{journal}{IEEE Transactions on Information Theory}
  \textbf{\bibinfo{volume}{52}}, \bibinfo{pages}{3456 } (\bibinfo{year}{2006}).

\bibitem[{\citenamefont{Ziv and Lempel}(1978)}]{lz78}
\bibinfo{author}{\bibfnamefont{J.}~\bibnamefont{Ziv}} \bibnamefont{and}
  \bibinfo{author}{\bibfnamefont{A.}~\bibnamefont{Lempel}},
  \bibinfo{journal}{IEEE T. Inform. Theory} \textbf{\bibinfo{volume}{24}},
  \bibinfo{pages}{530} (\bibinfo{year}{1978}).

\bibitem[{\citenamefont{Coutinho et~al.}(2010)\citenamefont{Coutinho, Fred, and
  Figueiredo}}]{coubio}
\bibinfo{author}{\bibfnamefont{D.~P.} \bibnamefont{Coutinho}},
  \bibinfo{author}{\bibfnamefont{A.~L.} \bibnamefont{Fred}}, \bibnamefont{and}
  \bibinfo{author}{\bibfnamefont{M.~A.} \bibnamefont{Figueiredo}},
  \bibinfo{journal}{Pattern Recognition, International Conference on}
  \textbf{\bibinfo{volume}{0}}, \bibinfo{pages}{3858} (\bibinfo{year}{2010}),
  ISSN \bibinfo{issn}{1051-4651}.

\bibitem[{\citenamefont{Kingman}(1969)}]{kingman}
\bibinfo{author}{\bibfnamefont{J.~F.~C.} \bibnamefont{Kingman}},
  \bibinfo{journal}{Journal of Applied Probability}
  \textbf{\bibinfo{volume}{6}}, \bibinfo{pages}{1} (\bibinfo{year}{1969}).

\bibitem[{\citenamefont{Ajdari and Prost}(1992)}]{prost}
\bibinfo{author}{\bibfnamefont{A.}~\bibnamefont{Ajdari}} \bibnamefont{and}
  \bibinfo{author}{\bibfnamefont{J.}~\bibnamefont{Prost}},
  \bibinfo{journal}{C.R. Acad. Sci. Paris II} \textbf{\bibinfo{volume}{315}},
  \bibinfo{pages}{1635} (\bibinfo{year}{1992}).

\bibitem[{mat()}]{matlab}
\emph{\bibinfo{title}{\uppercase{MATLAB R2011}b documentation (curve fitting
  toolbox)}},
  \bibinfo{howpublished}{\url{http://www.mathworks.es/help/toolbox/curvefit/}}.

\bibitem[{\citenamefont{Lacoste and Mallick}(2009)}]{lacoste}
\bibinfo{author}{\bibfnamefont{D.}~\bibnamefont{Lacoste}} \bibnamefont{and}
  \bibinfo{author}{\bibfnamefont{K.}~\bibnamefont{Mallick}},
  \bibinfo{journal}{Phys. Rev. E} \textbf{\bibinfo{volume}{80}},
  \bibinfo{pages}{021923} (\bibinfo{year}{2009}).

\bibitem[{\citenamefont{van Hemmen and Palmer}(1979)}]{Hemmen}
\bibinfo{author}{\bibfnamefont{J.}~\bibnamefont{van Hemmen}} \bibnamefont{and}
  \bibinfo{author}{\bibfnamefont{R.}~\bibnamefont{Palmer}},
  \bibinfo{journal}{J. Phys. A: Math. Gen.} \textbf{\bibinfo{volume}{12}},
  \bibinfo{pages}{563} (\bibinfo{year}{1979}).

\end{thebibliography}

\end{document}